\documentclass[aps,prd,superscriptaddress,preprint]{revtex4-1}

\usepackage{amsmath,amssymb,bm}
\usepackage{graphicx,graphics,color}
\usepackage[colorlinks=true,linkcolor=blue,bookmarks=false]{hyperref}
\usepackage{epsfig}
\usepackage{latexsym}
\usepackage{rotating}
\usepackage{subfigure}
\usepackage{centernot}
\usepackage{longtable}
\usepackage{hhline,multirow,tabularx}
\usepackage{float}
\usepackage{mathrsfs}
\usepackage{xr}
\usepackage{slashed}
\usepackage{adjustbox}

\newcommand{\RNum}[1]{\uppercase\expandafter{\romannumeral #1\relax}}

\newcommand{\yb}{\bar{y}}

\begin{document}

\preprint{JLAB-THY-20-3238, ADP-20-24/T1134}

\title{Strange quark helicity in the proton from chiral effective theory}

\author{X. G. Wang}
\affiliation{CoEPP and CSSM, Department of Physics, University of Adelaide, Adelaide SA 5005, Australia}
\author{Chueng-Ryong Ji}
\affiliation{Department of Physics, North Carolina State University, Raleigh, North Carolina 27695, USA}
\author{W. Melnitchouk}
\affiliation{Jefferson Lab, Newport News, Virginia 23606, USA}
\author{Y. Salamu}
\affiliation{Institute of High Energy Physics, CAS, Beijing 100049, China}
\affiliation{School of Physical Sciences, University of Chinese Academy	of Sciences, Beijing 100049, China}
\author{A. W. Thomas}
\affiliation{CoEPP and CSSM, Department of Physics, University of Adelaide, Adelaide SA 5005, Australia}
\author{P. Wang}
\affiliation{Institute of High Energy Physics, CAS, Beijing 100049, China}
\affiliation{Theoretical Physics Center for Science Facilities, CAS, Beijing 100049, China}

\begin{abstract}
We compute the helicity-dependent strange quark distribution in the proton in the framework of chiral effective theory.
Starting from the most general chiral SU(3) Lagrangian that respects Lorentz and gauge invariance, we derive the complete set of hadronic splitting functions at the one meson loop level, including the octet and decuplet rainbow, tadpole, Kroll-Ruderman and octet-decuplet transition configurations.
By matching hadronic and quark level operators, we obtain generalized convolution formulas for the quark distributions in the proton in terms of hadronic splitting functions and quark distributions in the hadronic configurations, and from these derive model-independent relations for the leading nonanalytic behavior of their moments.
Within the limits of parameters of the Pauli-Villars regulators derived from inclusive hyperon production, we find that the polarized strange quark distribution is rather small and mostly negative.
\end{abstract}

\date{\today}
\maketitle

\section{Introduction}
\label{sec:intro}

In 1987 the measurement by the European Muon Collaboration of the spin-dependent $g_1$ structure function of the proton led to the surprising conclusion that the sum of quark spins constituted a very small fraction of the spin of the proton~\cite{EMC:1988}.
The early polarized deep-inelastic scattering (DIS) measurements also suggested that a large fraction of the proton's spin may be carried by strange quarks~\cite{EMC:1989}, in stark contrast with simple quark model expectations (see Ref.~\cite{Aidala12} for a review).
Subsequent polarized DIS experiments with increasing precision and kinematic reach have been performed at 
    SLAC~\cite{SLAC-E142, SLAC-E143, SLAC-E154, SLAC-E155p, SLAC-E155d, SLAC-E155_A2pd, SLAC-E155x},
    HERMES~\cite{HERMES97, HERMES:2007, HERMES12},
    SMC~\cite{SMC98, SMC99},
    COMPASS~\cite{COMPASS:2007, COMPASSdis10} and
    Jefferson Lab~\cite{E99-117, CLAS:2006, eg1b-p-Prok, eg1-dvcs, eg1b-d, eg1b-p, E06-014_A1, E06-014_d2, E01-012, Armstrong:2018xgk},
and have provided a richer picture of the spin decomposition of the proton.

Data from these and other polarized high-energy scattering processes, such as jet and $W$ boson production in polarized $pp$ collisions at RHIC~\cite{STAR-W, PHENIX-W, STAR_jet19}, have been utilized in global QCD analyses of spin-dependent parton distribution functions (PDFs) by a number of groups~\cite{DSSV09, DSSV14, AAC09, BB10, LSS10, LSS11, LSS15, KTA17, NNPDF:2014, JAM15, JAM16, JAM17}.
The latest results from the JAM Collaboration's simultaneous analysis~\cite{JAM17} of helicity PDFs and fragmentation functions give a fraction $\Delta\Sigma = 0.36 \pm 0.09$ of the proton's spin carried by quarks and antiquarks at a scale of $Q^2=1$~GeV$^2$.
Parallel efforts from lattice QCD have also been made on calculations of moments of PDFs through the matrix elements of appropriate quark and gluon local operators within nucleon states~\cite{PNDME:2018, ETMC:2017, Engelhardt:2012, Gong:2017, Alexandrou:2020sml}, and more recently first studies have been explored of the feasibility of extracting information on the dependence of PDFs on the parton momentum fraction $x$ from quasi-PDF and pseudo-PDF lattice calculations~\cite{Alexandrou:2019lfo, Joo:2019jct}.

Among the three light quark flavors, the contribution to the proton spin from the strange quark is the least well determined, and phenomenological studies often rely on assumptions such as SU(3) flavor symmetry and equivalence of the strange and antistrange polarizations, $\Delta s = \Delta \bar{s}$, to simplify the analyses.
In many of the studies which have made these assumptions the strange quark polarization has typically been found to be in the vicinity of
    $\Delta s^+ \equiv \Delta s + \Delta \bar s \approx -0.1$.
Recent direct lattice simulations of disconnected loop contributions have yielded slightly smaller magnitudes for the strange quark polarization,
    $\Delta s^+_{\rm latt} = -0.046(8)$~\cite{Alexandrou:2020sml},
while an analysis of the spin problem taking into account the angular momentum carried by the meson cloud~\cite{Thomas:2008ga, Schreiber:1988uw, Myhrer:2007cf}, suggests a value of order $-0.01$~\cite{Bass:2009ed, Yamaguchi:1989sx}.
The recent JAM global QCD analysis, which used inclusive and semi-inclusive DIS data in order to relax the SU(3) symmetry constraint, also supports a smaller magnitude for the strange polarization,
    $\Delta s^+_{\rm JAM} = -0.03(10)$~\cite{JAM17}
at a scale of $Q^2=1$~GeV$^2$, but with a larger uncertainty.
A review of the status and results from the global QCD analysis and lattice QCD communities can be found in Ref.~\cite{Lin:2018}.
In an interesting recent analysis, the role of polarized nucleon strangeness in core-collapse supernova evolution was explored by Hobbs~{\it et~al.}~\cite{Hobbs:2016xlg}.

It was shown recently by de~Florian and Vogelsang \cite{Vogelsang:2019} that a nonzero integrated asymmetry between $\Delta s$ and $\Delta \bar s$ can arise from perturbative QCD evolution at three-loop order.
The effect was found to be small, however, with the difference
    $\Delta s - \Delta \bar s$
predicted to be negative and around 1\% of the sum
    $\Delta s + \Delta \bar s$.
This is in contrast to the unpolarized case, where the total number of strange and antistrange quarks must be equal, even though the shape of their momentum fraction distributions in $x$ need not be the same at three loops~\cite{Catani03}.

On the other hand, meson cloud models, in which the proton's strangeness content is generated by fluctuations to kaon-hyperon states such as $p \to \Lambda K^+$, naturally predict zero polarization for antistrange quarks.
In the limit in which the kaon mass is much smaller than the baryon masses, the $P$-wave nature of the kaon emission would require the $\Lambda$ to be polarized in the opposite direction to the proton.
Since in a nonrelativistic quark model picture the strange quark carries all of the spin of the $\Lambda$, the expectation would be for the strange quark polarization to be negative.
On the other hand, inclusion of relativistic effects~\cite{Malheiro97, Malheiro99}, as well as Fock states with higher-mass hyperons and $K^*$ mesons~\cite{Holtmann96, Zamani:2001, Cao:2003}, can significantly affect the shape and even the sign of the $\Delta s$ distribution.

A more systematic approach to computing the effects of pseudoscalar meson loops lies in the framework of chiral effective field theory, which establishes a more direct connection between the meson cloud of the nucleon and the underlying QCD theory.
This methodology has been applied recently in studies of the unpolarized light quark asymmetry
    $\bar d - \bar u$
and the strange--antistrange asymmetry
    $s - \bar s$
in the proton, using both local~\cite{Salamu:2018, plb-2016, prd-2016} and nonlocal~\cite{Salamu:2019-1,Salamu:2019-2} formulations.
Here, we extend our previous analysis~\cite{prd-2016} of the chiral loop contributions to the nonperturbative strange quark PDF to the polarized sector.
We work within the local formulation of the chiral effective theory, using Pauli-Villars to regularize the integrals and consider both the SU(3) octet and decuplet hadronic states.

In Sec.~\ref{sec:lagrangian}, we begin by presenting the lowest order meson-baryon chiral effective Lagrangian, consistent with Lorentz and gauge invariance.
The convolution formalism for the nucleon PDFs in the framework of chiral effective theory is discussed in Sec.~\ref{sec:formalism}, including the effective twist-2 operators relevant for the spin-dependent distributions.
Hadronic splitting functions are derived in Sec.~\ref{sec:fy}, including for the octet and decuplet rainbow diagrams, Kroll-Ruderman, tadpole, and octet-decuplet transition contributions, and from these the model-independent leading nonanalytic (LNA) behavior of the loop contributions to the moments of the PDFs is deduced in Sec.~\ref{sec:LNA}.
The regularization procedures dealing with the divergent loop integrals are discussed in Sec.~\ref{sec:regularization}, and the detailed numerical results for the polarized strange quark distributions in the proton are shown in Sec.~\ref{sec:results}.
Finally, we summarize our analysis and discuss future possible extensions of this work in Sec.~\ref{sec:conclusion}.
In Appendix~\ref{sec:fT}, we present some details about the derivation of the decuplet rainbow splitting function and the octet-decuplet splitting function. 

\section{Effective Lagrangian}
\label{sec:lagrangian}

In this section we review the basic effective chiral SU(3) Lagrangian describing the relativistic interactions of pseudoscalar mesons ($\phi$) and SU(3) octet ($B$) and decuplet ($T$) baryons~\cite{Jenkins:1991, Bernard:2008, Hacker:2005}.
To lowest order, this can be written as
\begin{eqnarray}
\label{eq:chiral-lagrangian}
\mathcal{L}
&=& i \left\langle \bar{B}\gamma^\mu [D_\mu,B] \right\rangle 
- \frac{1}{2} D \left\langle \bar{B}\gamma^\mu\gamma_5 \{u_\mu,B\} \right\rangle 
- \frac{1}{2} F \left\langle \bar{B}\gamma^\mu\gamma_5  [u_\mu,B]  \right\rangle
            \nonumber\\
& &
- \frac{1}{2} \mathcal{C}
    \Big[ \overline{T}_\mu \Theta^{\mu\nu} u_\nu B 
        + \bar{B} u_\mu \Theta^{\mu\nu} T_\nu
    \Big]
- \frac{1}{2} \mathcal{H}\, \overline{T}_\nu \gamma^\mu \gamma_5 u_\mu T^\nu,
\end{eqnarray}
where $D$ and $F$ are the meson--octet baryon coupling constants, and
$\mathcal{C}$ and $\mathcal{H}$ are the meson--octet--decuplet and meson--decuplet--decuplet baryon couplings, respectively.
In the meson sector the operator $u_\mu$ is defined as
\begin{equation}
u_\mu = i \left( u^\dag\partial_\mu u - u \partial_\mu u^\dag\right),
\end{equation}
with $u$ given in terms of the pseudoscalar fields $\phi$,
\begin{equation}
u = \exp\left(\frac{i\phi}{\sqrt{2} f_\phi}\right),
\end{equation}
and $f_\phi$ is the pseudoscalar meson decay constant.
The pseudoscalar pion, kaon and $\eta$ meson fields can be collected in the matrix $\phi$,
\begin{equation}
\phi
= \left(
  \begin{array}{ccc}
    \frac{1}{\sqrt{2}}\pi^0+\frac{1}{\sqrt{6}}\eta & \pi^+ & K^+ \\
    \pi^- & -\frac{1}{\sqrt{2}}\pi^0+\frac{1}{\sqrt{6}}\eta & K^0 \\
    K^- & \overline{K}^0 & -\frac{2}{\sqrt{6}}\eta \\
  \end{array}
\right).
\end{equation}
The covariant derivative $D^\mu$ in Eq.~(\ref{eq:chiral-lagrangian}) is defined by
\begin{equation}
[D_\mu,B] = \partial_\mu B + [\Gamma_\mu,B],
\end{equation}
where $\Gamma^{\mu}$ is the link operator,
\begin{equation}
\Gamma_\mu = \frac{1}{2} [u^{\dag}, \partial_{\mu}u].
\end{equation}
The SU(3) octet baryon fields $B$ are given by
\begin{equation}
B = \left(
          \begin{array}{ccc}
            \frac{1}{\sqrt{2}}\Sigma^0+\frac{1}{\sqrt{6}}\Lambda & \Sigma^+ & p \\
            \Sigma^- & -\frac{1}{\sqrt{2}}\Sigma^0+\frac{1}{\sqrt{6}}\Lambda & n \\
            \Xi^- & \Xi^0 & -\frac{2}{\sqrt{6}}\Lambda \\
          \end{array}
        \right),
\end{equation}
while the decuplet baryons may be included by way of a Rarita-Schwinger field, represented by the tensor $T^{ijk}$,
\begin{equation}
T = \frac{1}{\sqrt{3}}
\small
\left\{
\left(
\begin{array}{ccc}
\sqrt3\, \Delta^{++} & \Delta^+              & \Sigma^{*+} \\
\Delta^+            & \Delta^0              & \frac{1}{\sqrt{2}} \Sigma^{*0} \\
\Sigma^{*+}         & \frac{1}{\sqrt{2}}\Sigma^{*0} & \Xi^{*0}\
\end{array}
\right)
,
\left(
\begin{array}{ccc}
\Delta^+            & \Delta^0          & \frac{1}{\sqrt{2}} \Sigma^{*0} \\
\Delta^0            & \sqrt3 \Delta^-  & \Sigma^{*-}                   \\
\frac{1}{\sqrt{2}} \Sigma^{*0}          & \Sigma^{*-} & \Xi^{*-}        \\
\end{array}
\right)
,
\left(
\begin{array}{ccc}
\Sigma^{*+}         & \frac{1}{\sqrt{2}} \Sigma^{*0}    & \Xi^{*0}  \\
\frac{1}{\sqrt{2}} \Sigma^{*0} & \Sigma^{*-}            & \Xi^{*-}  \\
\Xi^{*0}            & \Xi^{*-}                          & \sqrt3\, \Omega^- \\
\end{array}
\right)
\right\}.
\end{equation}
\normalsize
The octet-decuplet transition tensor operator $\Theta^{\mu\nu}$ is defined as
\begin{equation}
\label{eq:Theta}
\Theta^{\mu\nu} 
= g^{\mu\nu} - \Big( Z + \frac{1}{2} \Big) \gamma^\mu \gamma^\nu,
\end{equation}
where $Z$ is the decuplet off-shell parameter.
To simplify the calculations, in this analysis we will choose $Z=1/2$~\cite{Scherer:2012xha}, although the physical results should be independent of the value of $Z$ chosen.
The octet--decuplet--meson interaction term in Eq.~(\ref{eq:chiral-lagrangian}) can be written explicitly in component form as~\cite{Labrenz:1996}
\begin{equation}
\overline{T}^{\, \mu} u_\mu B
= (\overline{T}^{\, \mu})_{ijk} (u_\mu)_{ii'} (B)_{jj'}\, \varepsilon_{i'j'k}\ .
\end{equation}

Expanding the effective Lagrangian (\ref{eq:chiral-lagrangian}) up to $\mathcal{O}\big((\phi/f_{\phi})^2\big)$, we can write this in more explicit fashion as a sum of specific meson--baryon interactions,
\begin{equation}
\mathcal{L}
    = \mathcal{L}_{\phi B B'}
    + \mathcal{L}_{\phi\phi B B} 
    + \mathcal{L}_{\phi B T}
    + \mathcal{L}_{\phi T T'}\ ,
\label{eq:Lexpand}
\end{equation}
where the first two terms, representing the meson--octet baryon interaction and the Weinberg-Tomozawa term, are given in Ref.~\cite{prd-2016}.
The third term involves the meson--octet--decuplet vertex and is given by
\begin{eqnarray}
\label{eq:phi-B-T}
\mathcal{L}_{\phi BT}
&=& \frac{{\cal C}}{\sqrt2 f_\phi} 
\left\{ 
- \frac{1}{\sqrt6} 
  \overline{\Sigma}^{*0}_\mu\, \Theta^{\mu\nu}\, \partial_\nu K^-\, p
+ \frac{1}{\sqrt3} 
  \overline{\Sigma}^{*+}_\mu\, \Theta^{\mu\nu} \partial_\nu \overline{K}^0\, p
+ \overline{\Delta}^{++}_\mu\, \Theta^{\mu\nu}\, \partial_\nu \pi^+\, p \right.                   \nonumber\\
&& \left. \hspace*{1.4cm}
- \sqrt{\frac23}\, 
  \overline{\Delta}^+_\mu\, \Theta^{\mu\nu}\, \partial_\nu \pi^0\, p 
- \frac{1}{\sqrt3} 
  \overline{\Delta}^0_\mu\, \Theta^{\mu\nu}\, \partial_\nu \pi^-\, p
+ {\rm h.c.}
\right\}.
\end{eqnarray}
The final term in Eq.~(\ref{eq:Lexpand}) involving the meson--decuplet--decuplet baryon vertices is not shown as it is not relevant to the matrix elements at the one-loop level when the initial and final states are both nucleons.

\section{Parton distributions in the nucleon}
\label{sec:formalism}

In this section, we derive the polarized PDFs in the nucleon within the convolution formalism by matching the spin-dependent twist-2 quark operators to hadronic operators with the same quantum numbers.
We identify the complete set of hadronic operators contributing to the polarized quark distributions, and relate the matching coefficients to the moments of PDFs in the hadronic configurations.

\subsection{Convolution formalism}
\label{ssec:convolution}

The $n{\rm th}$ Mellin moment of the spin-dependent quark distribution $\Delta q(x)$ is defined as
\begin{equation}
\label{eq:mom_def}
\langle x^{n-1} \rangle_{\Delta q}
\equiv \int_{-1}^1 dx\, x^{n-1} \Delta q(x)
= \int_0^1 dx\, x^{n-1}
  \Big( \Delta q(x) + (-1)^{n-1} \Delta\bar{q}(x) \Big),
\end{equation}
where we have used the crossing symmetry relation
    $\Delta q(-x) = + \Delta \bar q(x)$
between the quark and antiquark distributions.
(Note that spin-averaged PDFs, in contrast, have the opposite crossing symmetry property~\cite{prd-2016}.)
From the operator product expansion these moments can be related to the matrix elements of local twist-2 operators
    ${\cal O}_{\Delta q}^{\mu_1 \cdots \mu_n}$ 
between nucleon states,
\begin{equation}
\langle N(p,s) | \mathcal{O}_{\Delta q}^{\mu_1 \cdots \mu_n} | N(p,s) \rangle
= 2 \langle x^{n-1} \rangle_{\Delta q}\, M\,
  s^{ \{ \mu_1 } p^{\mu_2} \cdots p^{ \mu_n\} },
\end{equation}
where $p^\mu$ is the four-momentum of the nucleon and $s^\mu$ its polarization vector, with $s^2=-1$,
and the braces $\{ \cdots \}$ represent total symmetrization of Lorentz indices.
The spin-dependent twist-2 operators are defined as
\begin{equation}\label{eq:O-delta-q}
\mathcal{O}^{\mu_1 \cdots \mu_n}_{\Delta q}
= i^{n-1} \bar{q} \gamma_5 \gamma^{ \{\mu_1 }
  \overleftrightarrow{D}^{\mu_2} \cdots
  \overleftrightarrow{D}^{\mu_n\} } q,
\end{equation}
with $\overleftrightarrow{D} = \frac{1}{2} \big( \overrightarrow{D} - \overleftarrow{D} \big)$.
In an effective field theory, these quark operators are matched to hadronic operators with the same quantum numbers (but not necessarily with the same twist)~\cite{Ji:2001},
\begin{equation}\label{eq:operator-match}
\mathcal{O}^{\mu_1\cdots \mu_n}_{\Delta q}
= \sum_h c^{(n)}_{\Delta q/h}\, \widetilde{\mathcal{O}}^{\mu_1 \cdots \mu_n}_h\, ,
\end{equation}
where the subscript $h$ labels different types of hadronic operators.
The $c$-number coefficients $c^{(n)}_{\Delta q/h}$ can be defined through the $n{\rm th}$ moments of the spin-dependent PDFs $\Delta q_h(x)$ in the hadronic configuration $h$,
\begin{equation}
c^{(n)}_{\Delta q/h}
\equiv \langle x^{n-1} \rangle_{\Delta q/h}
= \int_0^1 dx\, x^{n-1} 
  \left[ \Delta q_h(x) + (-1)^{n-1} \Delta\bar{q}_h(x) \right].
\end{equation}
Matrix elements of the hadronic operators $\widetilde{\mathcal{O}}^{\mu_1 \cdots \mu_n}_h$ are used to define the moments of the hadronic splitting functions $\Delta f_h$ by taking the ``$+$'' components of the Lorentz indices,
\begin{equation}
\label{eq:Delta_fj}
\int_{-1}^1 dy\, y^{n-1} \Delta f_h(y)
= \frac{1}{2M s^+ (p^+)^{n-1}} \langle N(p,s) | \widetilde{\mathcal{O}}^{+\, \cdots\, +}_h | N(p,s) \rangle.
\end{equation}
In analogy with the unpolarized case~\cite{prd-2016}, the operator relation in Eq.~(\ref{eq:operator-match}) then gives rise to a convolution form for the spin-dependent PDFs in the nucleon,
\begin{equation}\label{eq:convolution-1}
\Delta q(x)
= \sum_h \big[\Delta f_h \otimes \Delta q_h^+\big](x)
\equiv \sum_h \int_0^1 dy\, \int_0^1 dz\, 
  \delta(x-yz)\, \Delta f_h(y)\, \Delta q_h^+(z),
\end{equation}
where $\Delta q_h^+ = \Delta q_h + \Delta \bar{q}_h$ is the spin-dependent quark distribution for quark flavor $q$ in the hadronic configuration $h$.
The convolution expression (\ref{eq:convolution-1}) is the basis for the calculation of the contributions to the quark helicity distributions from the chiral loop corrections generated from the Lagrangian (\ref{eq:chiral-lagrangian}).

\subsection{Twist-2 operators}

The spin-dependent quark operators in Eq.~(\ref{eq:O-delta-q}) can be matched to hadronic operators derived from the lowest order Lagrangian in Eq.~(\ref{eq:Lexpand})~\cite{Salamu:2019-2, Shanahan:2013},
\begin{eqnarray}\label{eq:O-delta-q-hadronic}
\mathcal{O}^{\mu_1 \cdots \mu_n}_{\Delta q}
&=& \Big[
    \bar{\alpha}^{(n)}
    \big( \overline{{\cal B}} \gamma^{\mu_1}\gamma_5 {\cal B} \lambda^q_+ \big)
 +  \bar{\beta}^{(n)}
    \big( \overline{{\cal B}} \gamma^{\mu_1}\gamma_5 \lambda^q_+ {\cal B} \big)
 +  \bar{\sigma}^{(n)}
    \big( \overline{{\cal B}} \gamma^{\mu_1}\gamma_5 {\cal B} \big) \mathrm{Tr}\lambda^q_+
    \Big]\, p^{\mu_2} \ldots p^{\mu_n}
\nonumber\\
&+& \Big[ 
    \alpha^{(n)}
    \big( \overline{{\cal B}} \gamma^{\mu_1} {\cal B} \lambda^q_- \big)
 +  \beta^{(n)}
    \big( \overline{{\cal B}} \gamma^{\mu_1} \lambda^q_- {\cal B} \big)
 +  \sigma^{(n)}
    \big( \overline{{\cal B}} \gamma^{\mu_1} {\cal B} \big) \mathrm{Tr}\lambda^q_-
    \Big]\, p^{\mu_2} \ldots p^{\mu_n}
\nonumber\\   
&+& \Big[ 
    \bar{\gamma}^{(n)}
    \big( \overline{T}^{\nu} \gamma^{\mu_1}\gamma_5 \lambda^q_+ T_\nu \big)
 -  \sqrt{\frac32}\, \bar{\omega}^{(n)}
    \big[ 
    \big( \overline{T}_\nu \Theta^{\nu\mu_1} \lambda^q_+ {\cal B} \big)
   +\big( \overline{{\cal B}} \lambda^q_+ \Theta^{\mu_1\nu} T_\nu \big)
    \big] 
    \Big]\, p^{\mu_2} \ldots p^{\mu_n}    \nonumber\\
&+& \mathrm{permutations}\ -\ \mathrm{Tr},
\end{eqnarray}
where the trace ``Tr'' here is over the Lorentz indices.
The {\it a priori} unknown coefficients
    $\{ \bar\alpha^{(n)}, \bar\beta^{(n)}, \bar\sigma^{(n)} \}$
and
    $\{ \alpha^{(n)}, \beta^{(n)}, \sigma^{(n)} \}$
correspond to the octet baryonic pseudovector and vector operators, respectively, while 
$\bar{\gamma}^{(n)}$ and $\bar{\omega}^{(n)}$ correspond to decuplet-decuplet and octet-decuplet transition operators, respectively.
Note that only those operators that contribute to matrix elements with initial and final nucleon states are listed in Eq.~(\ref{eq:O-delta-q-hadronic}).

Writing the spin-1/2 octet baryon operator ${\cal B}$ in a three-index tensor representation, one can relate this to the octet baryon field matrix $B$ by
\begin{eqnarray}
\label{eq:Octet_representation}
{\cal B}_{ijk}
&=& \frac{1}{\sqrt6}
    \big( \epsilon_{ijk'} B^{k'}_k + \epsilon_{ikk'} B^{k'}_j
    \big),
\end{eqnarray}
with the corresponding conjugate representation giving
\begin{eqnarray}
\label{eq:Octet_conj_representation}
\overline{\cal B}_{kji}
&=& \frac{1}{\sqrt6}
    \big( \epsilon_{ijk'} \bar{B}^{k'}_k + \epsilon_{ikk'} \bar{B}^{k'}_j
    \big),
\end{eqnarray}
where $\epsilon_{ijk}$ is the antisymmetric tensor. 
In Eq.~(\ref{eq:O-delta-q-hadronic}) the flavor operator $\lambda^q_\pm$ is defined as
\begin{eqnarray}
\lambda^q_{\pm}
&=& \frac12
    \left( u\bar{\lambda}^q u^{\dag} \pm u^{\dag} \bar{\lambda}^q u
    \right),
\end{eqnarray}
with $\bar{\lambda}^q = {\rm diag}(\delta_{qu}, \delta_{qd}, \delta_{qs})$ being diagonal $3 \times 3$ matrices.
Expanding $\lambda^q_\pm$ up to ${\cal O}(\phi^2)$, one has
\begin{subequations}
\label{eq:lambdaq}
\begin{eqnarray}
\lambda^q_+
&=& \bar{\lambda}^q
 +  \frac{1}{4 f_\phi^2}
    \Big( 2\phi \bar{\lambda}^q \phi - \phi^2 \bar{\lambda}^q
	- \bar{\lambda}^q \phi^2
    \Big)
 +  {\cal O}\left(\phi^4\right),		\\
\lambda^q_-
&=& \frac{i}{\sqrt{2} f_\phi}
    \Big( \phi \bar{\lambda}^q -\bar{\lambda}^q \phi
    \Big) 
 +  {\cal O}\left(\phi^3\right).
\end{eqnarray}
\end{subequations}
Finally, the combinations of operators 
    $\big( \bar{\mathcal{B}} \cdots \mathcal{B} \big)$, 
    $\big( \overline{T}_\mu A T_\nu \big)$
and $\big( \overline{T}_\mu A \mathcal{B} \big)$ in Eq.~(\ref{eq:O-delta-q-hadronic}) involving the three-index tensors are given by~\cite{Labrenz:1996}
\begin{subequations}
\begin{eqnarray}
(\overline{\cal B} {\cal B})
&=& \mathrm{Tr}\big[\bar{B}B\big],			\\
(\overline{\cal B} {\cal B} A)
&=& \frac{2}{3}\mathrm{Tr}\big[\bar{B} A B\big]
 +  \frac{1}{6}\mathrm{Tr}\big[\bar{B}B\big] \mathrm{Tr}\big[A\big]
 -  \frac{1}{6}\mathrm{Tr}\big[\bar{B} B A\big],	\\
(\overline{\cal B} A {\cal B})
&=& -\frac{1}{3}\mathrm{Tr}\big[\bar{B} A B\big]
 +  \frac{2}{3}\mathrm{Tr}\big[\bar{B}B\big] \mathrm{Tr}\big[A\big]
 -  \frac{2}{3}\mathrm{Tr}\big[\bar{B} B A\big],    
\end{eqnarray}
\end{subequations}
and
\begin{subequations}
\begin{eqnarray}
\big( \overline{T}_\mu A T_{\nu} \big)
&=& \overline{T}^{kji}_\mu\, A^{il}\, T_\nu^{ljk},  \\
\big( \overline{T}_{\mu} A \mathcal{B} \big)
&=& -\sqrt{\frac23} \overline{T}^{ijk}_\mu A^{i i'}\, \mathcal{B}^{j j'} \epsilon^{i' j' k}.
\end{eqnarray}
\end{subequations}
With these relations we can write the hadronic operators explicitly for each of the spin-dependent $u$, $d$ and $s$ quark distributions as
\begin{eqnarray}\label{eq:Ou}
\mathcal{O}^{\mu_1 \cdots \mu_n}_{\Delta u} 
&=&
\Big( \frac56 \bar{\alpha}^{(n)}
    + \frac13 \bar{\beta}^{(n)}
    + \bar{\sigma}^{(n)}
\Big) \widetilde{\mathcal{O}}^{\mu_1 \cdots \mu_n}_{p}
+
\Big( \frac16 \bar{\alpha}^{(n)}
    + \frac23 \bar{\beta}^{(n)}
    + \bar{\sigma}^{(n)}
\Big) \widetilde{\mathcal{O}}^{\mu_1 \cdots \mu_n}_{n}  		\nonumber\\
&+& 
\Big( \frac16 \bar{\alpha}^{(n)}
    + \frac23 \bar{\beta}^{(n)}
    + \bar{\sigma}^{(n)}
\Big) \widetilde{\mathcal{O}}^{\mu_1 \cdots \mu_n}_{\Xi^0} 
+
\Big( \frac14 \bar{\alpha}^{(n)}
    + \frac12 \bar{\beta}^{(n)}
    + \bar{\sigma}^{(n)}
\Big) \widetilde{\mathcal{O}}^{\mu_1 \cdots \mu_n}_{\Lambda}	\nonumber\\
&+& 
\Big( \frac{5}{12} \bar{\alpha}^{(n)}
    + \frac16 \bar{\beta}^{(n)}
    + \bar{\sigma}^{(n)}
\Big) \widetilde{\mathcal{O}}^{\mu_1 \cdots \mu_n}_{\Sigma^0} 
+
\Big( \frac56 \bar{\alpha}^{(n)}
    + \frac13 \bar{\beta}^{(n)}
    + \bar{\sigma}^{(n)}
\Big) \widetilde{\mathcal{O}}^{\mu_1 \cdots \mu_n}_{\Sigma^+} 	\nonumber\\
&+& 
\bar{\sigma}^{(n)}
\Big( \widetilde{\mathcal{O}}^{\mu_1 \cdots \mu_n}_{\Sigma^-}
      + \widetilde{\mathcal{O}}^{\mu_1 \cdots \mu_n}_{\Xi^-}
\Big)
+
\frac{1}{4\sqrt3}
\Big( \bar{\alpha}^{(n)} - 2 \bar{\beta}^{(n)} \Big)
\Big( \widetilde{\mathcal{O}}^{\mu_1 \cdots \mu_n}_{\Lambda \Sigma^0} 
    + \widetilde{\mathcal{O}}^{\mu_1 \cdots \mu_n}_{\Sigma^0 \Lambda}
\Big)	                                         \nonumber\\
&+&
\frac{1}{12} 
\Big[
  \big(-4 \bar\alpha^{(n)} + 2 \bar\beta^{(n)} \big)
  \widetilde{\mathcal{O}}^{\mu_1 \cdots \mu_n}_{\bar{p}p\pi^+ \pi^-}
- \big(5 \bar\alpha^{(n)} + 2 \bar\beta^{(n)} \big)
  \widetilde{\mathcal{O}}^{\mu_1 \cdots \mu_n}_{\bar{p}pK^+ K^-}
\nonumber\\
& & \hspace{0.9cm}
+ \big( 4 \bar\alpha^{(n)} - 2 \bar\beta^{(n)} \big)
  \widetilde{\mathcal{O}}^{\mu_1 \cdots \mu_n}_{\bar{n}n\pi^+ \pi^-}
- \big( \bar\alpha^{(n)} + 4 \bar\beta^{(n)} \big)
  \widetilde{\mathcal{O}}^{\mu_1 \cdots \mu_n}_{\bar{n}nK^+ K^-}
\Big]                   				\nonumber\\ 
&+& 
  \frac{\big( 2 \alpha^{(n)} - \beta^{(n)} \big)}{3\sqrt2}\,
  \widetilde{\mathcal{O}}^{\mu_1 \cdots \mu_n}_{n p \pi^-}
- \frac{\sqrt3 \alpha^{(n)}}{4}\,
  \widetilde{\mathcal{O}}^{\mu_1 \cdots \mu_n}_{\Lambda p K^-}
- \frac{\big( \alpha^{(n)} + 4 \beta^{(n)} \big)}{12} 
  \Big( \widetilde{\mathcal{O}}^{\mu_1 \cdots \mu_n}_{\Sigma^0 p K^-}
  + \sqrt2\, \widetilde{\mathcal{O}}^{\mu_1 \cdots \mu_n}_{\Sigma^- n K^-}
  \Big) \nonumber\\
&+&
\frac13 \bar{\gamma}^{(n)} 
\Big[
  3 \widetilde{\mathcal{O}}^{\mu_1 \cdots \mu_n}_{\Delta^{++}}
+ 2 \widetilde{\mathcal{O}}^{\mu_1 \cdots \mu_n}_{\Delta^{+}}
+   \widetilde{\mathcal{O}}^{\mu_1 \cdots \mu_n}_{\Delta^{0}}  
+ 2 \widetilde{\mathcal{O}}^{\mu_1 \cdots \mu_n}_{\Sigma^{*+}}
+   \widetilde{\mathcal{O}}^{\mu_1 \cdots \mu_n}_{\Sigma^{*0}} 
+   \widetilde{\mathcal{O}}^{\mu_1 \cdots \mu_n}_{\Xi^{*0}}
\Big]
\nonumber\\
&+&
\frac{1}{\sqrt3}\, \bar{\omega}^{(n)}
\Big[
  \widetilde{\mathcal{O}}^{\mu_1 \cdots \mu_n}_{\Delta^+ p}
+ \widetilde{\mathcal{O}}^{\mu_1 \cdots \mu_n}_{\Delta^0 n}
- \widetilde{\mathcal{O}}^{\mu_1 \cdots \mu_n}_{\Sigma^{*+}\Sigma^+}
+ \frac12 \widetilde{\mathcal{O}}^{\mu_1 \cdots \mu_n}_{\Sigma^{*0}\Sigma^0} 
- \frac{\sqrt3}{2} \widetilde{\mathcal{O}}^{\mu_1 \cdots \mu_n}_{\Sigma^{*0}\Lambda}
- \widetilde{\mathcal{O}}^{\mu_1 \cdots \mu_n}_{\Xi^{*0}\Xi^0}
\Big], \nonumber\\
\end{eqnarray}
\begin{eqnarray}\label{eq:Od}
\mathcal{O}^{\mu_1 \cdots \mu_n}_{\Delta d} 
&=&
\Big( \frac16 \bar{\alpha}^{(n)}
    + \frac23 \bar{\beta}^{(n)}
    + \bar{\sigma}^{(n)}
\Big) \widetilde{\mathcal{O}}^{\mu_1 \cdots \mu_n}_{p}
+
\Big( \frac56 \bar{\alpha}^{(n)}
    + \frac13 \bar{\beta}^{(n)}
    + \bar{\sigma}^{(n)}
\Big) \widetilde{\mathcal{O}}^{\mu_1 \cdots \mu_n}_{n}		\nonumber\\
&+& 
\Big( \frac16 \bar{\alpha}^{(n)}
    + \frac23 \bar{\beta}^{(n)}
    + \bar{\sigma}^{(n)}
\Big) \widetilde{\mathcal{O}}^{\mu_1 \cdots \mu_n}_{\Xi^-} 
+
\Big( \frac14 \bar{\alpha}^{(n)}
    + \frac12 \bar{\beta}^{(n)}
    + \bar{\sigma}^{(n)}
\Big) \widetilde{\mathcal{O}}^{\mu_1 \cdots \mu_n}_{\Lambda}	\nonumber\\
&+&
\Big( \frac{5}{12} \bar{\alpha}^{(n)}
    + \frac16 \bar{\beta}^{(n)}
    + \bar{\sigma}^{(n)}
\Big) \widetilde{\mathcal{O}}^{\mu_1 \cdots \mu_n}_{\Sigma^0} 
+
\Big( \frac56 \bar{\alpha}^{(n)}
    + \frac13 \bar{\beta}^{(n)}
    + \bar{\sigma}^{(n)}
\Big) \widetilde{\mathcal{O}}^{\mu_1 \cdots \mu_n}_{\Sigma^-} 	\nonumber\\
&+&
\bar{\sigma}^{(n)}
\Big( \widetilde{\mathcal{O}}^{\mu_1 \cdots \mu_n}_{\Sigma^+}
    + \widetilde{\mathcal{O}}^{\mu_1 \cdots \mu_n}_{\Xi^0}
\Big)
-
\frac{1}{4\sqrt{3}}
\Big( \bar{\alpha}^{(n)} - 2 \bar{\beta}^{(n)} \Big)
\Big( \widetilde{\mathcal{O}}^{\mu_1 \cdots \mu_n}_{\Lambda \Sigma^0} 
    + \widetilde{\mathcal{O}}^{\mu_1 \cdots \mu_n}_{\Sigma^0 \Lambda}
\Big)							\nonumber\\ 
&+&
\frac{1}{12}
\Big[
  \big( 4 \bar{\alpha}^{(n)} - 2 \bar{\beta}^{(n)} \big)
  \widetilde{\mathcal{O}}^{\mu_1 \cdots \mu_n}_{\bar{p}p\pi^+ \pi^-}
- \big( \bar{\alpha}^{(n)} + 4 \bar{\beta}^{(n)} \big)
  \widetilde{\mathcal{O}}^{\mu_1 \cdots \mu_n}_{\bar{p}pK^0 \bar{K}^0}
\nonumber\\
&& \hspace*{0.4cm}
- \big( 4 \bar{\alpha}^{(n)} - 2 \bar{\beta}^{(n)} \big)
  \widetilde{\mathcal{O}}^{\mu_1 \cdots \mu_n}_{\bar{n}n\pi^+ \pi^-}
- \big( 5 \bar{\alpha}^{(n)} + 2 \bar{\beta}^{(n)} \big)
  \widetilde{\mathcal{O}}^{\mu_1 \cdots \mu_n}_{\bar{n}nK^0 \bar{K}^0}
\Big]                   				\nonumber\\ 
&-& 
  \frac{\big( 2 \alpha^{(n)} - \beta^{(n)} \big)}{3\sqrt2}\,
  \widetilde{\mathcal{O}}^{\mu_1 \cdots \mu_n}_{n p \pi^-}
- \frac{\sqrt{3}}{4} \alpha^{(n)}\,
  \widetilde{\mathcal{O}}^{\mu_1 \cdots \mu_n}_{\Lambda n \bar{K}^0}
+ \frac{\big( \alpha^{(n)} + 4 \beta^{(n)} \big)}{12} 
  \Big(
  \widetilde{\mathcal{O}}^{\mu_1 \cdots \mu_n}_{\Sigma^0 n \bar{K}^0}
- \sqrt2 \widetilde{\mathcal{O}}^{\mu_1 \cdots \mu_n}_{\Sigma^+ p \bar{K}^0}
  \Big) \nonumber\\
&+&
\frac13 \bar{\gamma}^{(n)} 
\Big[
  3 \widetilde{\mathcal{O}}^{\mu_1 \cdots \mu_n}_{\Delta^{-}}
+ 2 \widetilde{\mathcal{O}}^{\mu_1 \cdots \mu_n}_{\Delta^{0}}
+   \widetilde{\mathcal{O}}^{\mu_1 \cdots \mu_n}_{\Delta^{+}}
+ 2 \widetilde{\mathcal{O}}^{\mu_1 \cdots \mu_n}_{\Sigma^{*-}} 
+   \widetilde{\mathcal{O}}^{\mu_1 \cdots \mu_n}_{\Sigma^{*0}}
+   \widetilde{\mathcal{O}}^{\mu_1 \cdots \mu_n}_{\Xi^{*-}}
\Big]
\nonumber\\
&+&
\frac{1}{\sqrt3} \bar{\omega}^{(n)} 
\Big[
  \widetilde{\mathcal{O}}^{\mu_1 \cdots \mu_n}_{\Delta^+ p}
+ \widetilde{\mathcal{O}}^{\mu_1 \cdots \mu_n}_{\Delta^0 n}
- \widetilde{\mathcal{O}}^{\mu_1 \cdots \mu_n}_{\Sigma^{*-}\Sigma^-}  
- \frac12 \widetilde{\mathcal{O}}^{\mu_1 \cdots \mu_n}_{\Sigma^{*0}\Sigma^0} 
- \frac{\sqrt3}{2} \widetilde{\mathcal{O}}^{\mu_1 \cdots \mu_n}_{\Sigma^{*0}\Lambda}
- \widetilde{\mathcal{O}}^{\mu_1 \cdots \mu_n}_{\Xi^{*-}\Xi^-}
\Big], \nonumber\\
\end{eqnarray}
\begin{eqnarray}\label{eq:Os} 
\mathcal{O}^{\mu_1 \cdots \mu_n}_{\Delta s}
&=& \Big( \frac12 \bar{\alpha}^{(n)} + \bar{\sigma}^{(n)}
    \Big)
    \widetilde{\mathcal{O}}^{\mu_1 \cdots \mu_n}_\Lambda       
 +  \Big( \frac16 \bar{\alpha}^{(n)}
        + \frac23 \bar{\beta}^{(n)}
        + \bar{\sigma}^{(n)}
    \Big)
    \Big( \widetilde{\mathcal{O}}^{\mu_1 \cdots \mu_n}_{\Sigma^+}
        + \widetilde{\mathcal{O}}^{\mu_1 \cdots \mu_n}_{\Sigma^0}
        + \widetilde{\mathcal{O}}^{\mu_1 \cdots \mu_n}_{\Sigma^-} 
    \Big)
    \nonumber\\
&+& \Big( \frac56 \bar{\alpha}^{(n)}
        + \frac13 \bar{\beta}^{(n)}
        + \bar{\sigma}^{(n)}
    \Big)
    \Big( \widetilde{\mathcal{O}}^{\mu_1 \cdots \mu_n}_{\Xi^-}
        + \widetilde{\mathcal{O}}^{\mu_1 \cdots \mu_n}_{\Xi^0}
    \Big)
 +  \bar{\sigma}^{(n)}
    \Big( \widetilde{\mathcal{O}}^{\mu_1\cdots \mu_n}_{p}
        + \widetilde{\mathcal{O}}^{\mu_1\cdots \mu_n}_{n}
    \Big)
    \nonumber\\
&+& \frac{1}{12}
\Big[
    \big( 5 \bar{\alpha}^{(n)} + 2 \bar{\beta}^{(n)} \big)
    \Big( \widetilde{\mathcal{O}}^{\mu_1 \cdots \mu_n}_{\bar{p}pK^+ K^-}
        + \widetilde{\mathcal{O}}^{\mu_1 \cdots \mu_n}_{\bar{n}nK^0 \bar{K}^0}
    \Big)
 +  \big( \bar{\alpha}^{(n)} + 4 \bar{\beta}^{(n)} \big)
    \Big( \widetilde{\mathcal{O}}^{\mu_1 \cdots \mu_n}_{\bar{p}pK^0\bar{K}^0}
        + \widetilde{\mathcal{O}}^{\mu_1 \cdots \mu_n}_{\bar{n}nK^+K^-}
    \Big)
\Big]    \nonumber\\
&+& \frac{\sqrt3}{4} \alpha^{(n)}
    \Big( \widetilde{\mathcal{O}}^{\mu_1\cdots \mu_n}_{\Lambda p K^-} 
        + \widetilde{\mathcal{O}}^{\mu_1\cdots \mu_n}_{\Lambda n \bar{K}^0}
    \Big)
 +  \frac{1}{12} \big( \alpha^{(n)} + 4 \beta^{(n)} \big)
    \Big( \widetilde{\mathcal{O}}^{\mu_1 \cdots \mu_n}_{\Sigma^0 p K^-}
        + \sqrt2 \widetilde{\mathcal{O}}^{\mu_1\cdots \mu_n}_{\Sigma^+ p \bar{K}^0}
    \Big)
    \nonumber\\
&-& \frac{1}{12} \big( \alpha^{(n)} + 4 \beta^{(n)} \big)
    \Big( \widetilde{\mathcal{O}}^{\mu_1 \cdots \mu_n}_{\Sigma^0 n \bar{K}^0}
    - \sqrt{2} \widetilde{\mathcal{O}}^{\mu_1 \cdots \mu_n}_{\Sigma^- n K^-}
    \Big)
    \nonumber\\
&+& \frac13 \bar{\gamma}^{(n)} 
\Big[
    \widetilde{\mathcal{O}}^{\mu_1 \cdots \mu_n}_{\Sigma^{*+}}
+   \widetilde{\mathcal{O}}^{\mu_1 \cdots \mu_n}_{\Sigma^{*0}}
+   \widetilde{\mathcal{O}}^{\mu_1 \cdots \mu_n}_{\Sigma^{*-}}
+ 2 \widetilde{\mathcal{O}}^{\mu_1 \cdots \mu_n}_{\Xi^{*-}}
+ 2 \widetilde{\mathcal{O}}^{\mu_1 \cdots \mu_n}_{\Xi^{*0}}
+ 3 \widetilde{\mathcal{O}}^{\mu_1 \cdots \mu_n}_{\Omega^{-}}
\Big]        \nonumber\\
&-& \frac{1}{\sqrt3}\, \bar{\omega}^{(n)} 
\Big[
  \widetilde{\mathcal{O}}^{\mu_1 \cdots \mu_n}_{\Sigma^{*+}\Sigma^+}
- \widetilde{\mathcal{O}}^{\mu_1 \cdots \mu_n}_{\Sigma^{*0}\Sigma^0}
- \widetilde{\mathcal{O}}^{\mu_1 \cdots \mu_n}_{\Sigma^{*-}\Sigma^-}
+ \widetilde{\mathcal{O}}^{\mu_1 \cdots \mu_n}_{\Xi^{*0}\Xi^0} 
- \widetilde{\mathcal{O}}^{\mu_1 \cdots \mu_n}_{\Xi^{*-}\Xi^-}
\Big].
\end{eqnarray}
The hadronic operators appearing in Eqs.~(\ref{eq:Ou})--(\ref{eq:Os}) are given by
\begin{subequations}
\label{eq:hadronic-operators}
\begin{eqnarray}
\widetilde{\mathcal{O}}^{\mu_1 \cdots \mu_n}_B
&=& \big( \bar{B} \gamma^{\mu_1} \gamma_5 B \big)\,
    p^{\mu_2} \ldots p^{\mu_n},     \\
\widetilde{\mathcal{O}}^{\mu_1 \cdots \mu_n}_{B'B}
&=& \big( \bar{B'} \gamma^{\mu_1} \gamma_5 B \big)\,
    p^{\mu_2} \ldots p^{\mu_n},     \\
\widetilde{\mathcal{O}}^{\mu_1 \cdots \mu_n}_{BB\phi\phi}
&=& \frac{1}{f^2_\phi} 
    \big( \bar{B} \gamma^{\mu_1} \gamma_5 B\, \bar{\phi} \phi \big)\,
    p^{\mu_2} \ldots p^{\mu_n},     \\
\widetilde{\mathcal{O}}^{\mu_1 \cdots \mu_n}_{B'B \phi}
&=& \frac{i}{f_\phi}
    \big( \bar{B'} \gamma^{\mu_1} B \phi - \bar{B} \gamma^{\mu_1} B' \bar{\phi} \big)\,
    p^{\mu_2} \ldots p^{\mu_n},
\end{eqnarray}
\end{subequations}
for octet baryon operators, and
\begin{subequations}
\label{eq:hadronic-operators-T}
\begin{eqnarray}
\widetilde{\mathcal{O}}^{\mu_1 \cdots \mu_n}_T
&=& \big( \overline{T}^\nu \gamma^{\mu_1} \gamma_5 T_\nu \big)\,
    p^{\mu_2} \ldots p^{\mu_n}, \\
\widetilde{\mathcal{O}}^{\mu_1 \cdots \mu_n}_{TB}
&=& \big( \overline{T}_\nu \Theta^{\nu\mu_1} B + \bar{B}\, \Theta^{\mu_1\nu} T_\nu \big)\,
    p^{\mu_2} \ldots p^{\mu_n},
\end{eqnarray}
\end{subequations}
for operators involving decuplet baryon fields.

\begin{figure}[t]
\vspace{0.5cm}
\center
\includegraphics[width=\columnwidth]{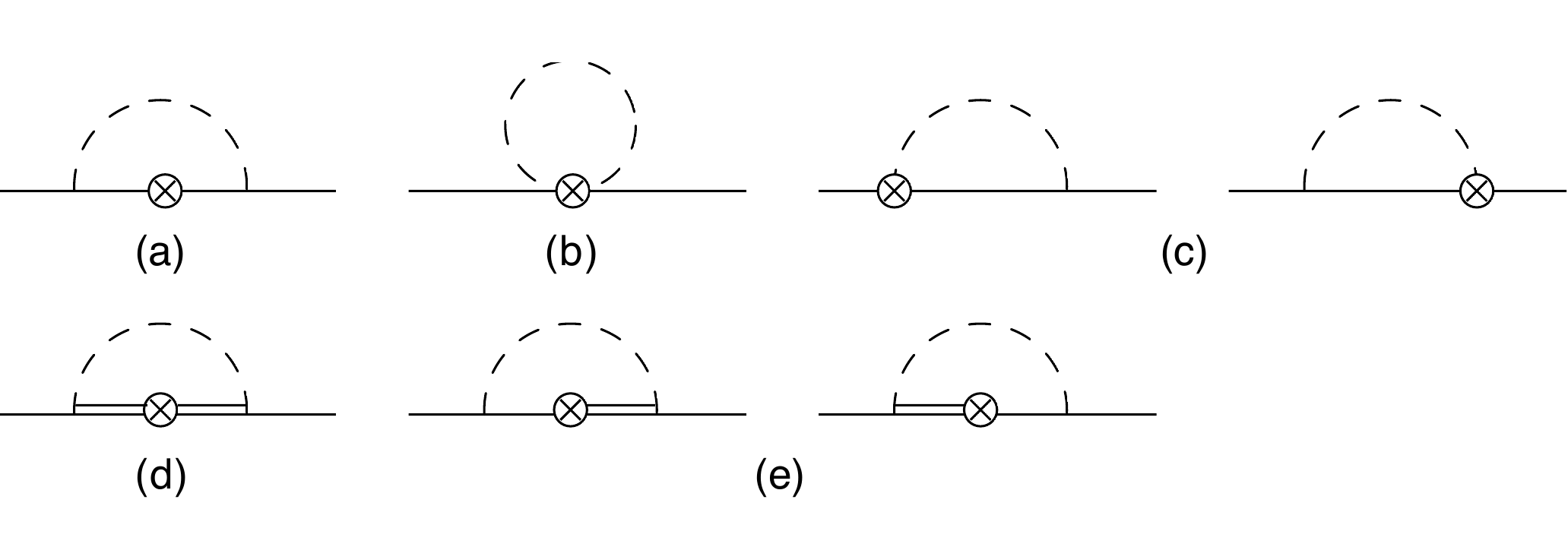}
\caption{One-loop contributions to the spin-dependent PDFs of the nucleon from
	{\bf (a)} octet rainbow,
	{\bf (b)} tadpole,
	{\bf (c)} Kroll-Ruderman,
	{\bf (d)} decuplet rainbow, and
	{\bf (e)} octet-decuplet transition 
diagrams. 
The octet baryons, decuplet baryons and pseudoscalar mesons are represented by the solid, double-solid and dashed lines, respectively, while the symbol $\otimes$ denotes insertion of the hadronic operators defined in Eqs.~(\ref{eq:Ou})--(\ref{eq:Os}).}
\label{fig:loop-octet-decuplet}
\end{figure}

In the present work we will focus on the polarized strange quark distributions in the proton, $\Delta s(x)$.
Correspondingly, the matrix elements of the hadronic operators give rise to the octet rainbow, tadpole, Kroll-Ruderman, decuplet rainbow, and octet-decuplet transition splitting functions, as illustrated by the diagrams in Fig.~\ref{fig:loop-octet-decuplet}.
The convolution representation (\ref{eq:convolution-1}) then gives the strange quark PDF in terms of the explicit hadronic configurations as
\begin{eqnarray}
\label{eq:convolution-2}
\Delta s(x)
&=& \sum_{B\phi}
    \Big( \Delta \bar{f}_{B\phi}^{(\rm rbw)} \otimes \Delta s_B
        + \Delta \bar{f}_{B\phi}^{(\rm KR)}  \otimes \Delta s_B^{(\rm KR)}
    \Big)
 +\ \sum_\phi \Delta \bar{f}_\phi^{(\rm tad)} \otimes \Delta s_\phi^{(\rm tad)}\nonumber\\
&+& \sum_{T \phi}
    \Delta \bar{f}_{T \phi}^{(\rm rbw)} \otimes \Delta s_{T}
    +\sum_{T B \phi} \Delta \bar{f}_{TB \phi} \otimes \Delta s_{T B}\ ,   
\end{eqnarray}
where for notational convenience we define the splitting functions
    $\bar{f}_j(y) \equiv f_j(\yb)$,
with \mbox{$\yb \equiv 1-y$} the baryon momentum fraction when the meson carries momentum fraction $y$.
For strange quarks the hadron labels span the
  mesons $\phi = K^0, K^+$;
  octet baryons $B = \Lambda, \Sigma^0, \Sigma^+$;
and
  decuplet baryons $T = \Sigma^{*0}, \Sigma^{*+}$.
The strange quark distributions in the various hadronic configurations include the strange quark PDFs in the octet and decuplet baryons, $\Delta s_B$ or $\Delta s_T$
    [Figs.~\ref{fig:loop-octet-decuplet}(a) and~\ref{fig:loop-octet-decuplet}(d)],
the transition decuplet-octet PDF, $\Delta s_{TB}$
    [Fig.~\ref{fig:loop-octet-decuplet}(e)],
the tadpole distributions, $\Delta s_\phi^{(\rm tad)}$
    [Fig.~\ref{fig:loop-octet-decuplet}(b)],
and the Kroll-Ruderman distributions, $\Delta s_B^{(\rm KR)}$
    [Fig.~\ref{fig:loop-octet-decuplet}(c)].
Note that while the convolution result in Eq.~(\ref{eq:convolution-2}) involves the $\Delta s_j^+$ distribution in the hadronic configuration, in our calculations we shall assume that all of the antiquarks reside in the pseudoscalar meson loops, so that the antiquark polarization is zero, $\Delta \bar{s}_j = 0$.
In the next section we discuss the calculation of these PDFs in more detail.

\newpage
\subsection{PDFs in hadronic configurations}
\label{ssec:hadronic-pdfs}

The spin-dependent strange quark distributions in the hadronic configurations as appearing in Eq.~(\ref{eq:convolution-2}) can be computed by relating their moments to the coefficients of the various terms in the twist-2 operator for the strange quark in Eq.~(\ref{eq:Os}).
Starting with the PDFs in the bare octet baryons, $\Delta s_B$ [Fig.~\ref{fig:loop-octet-decuplet}(a)], the moments can be expressed in terms of the coefficients $\bar\alpha^{(n)}$, $\bar\beta^{(n)}$ and $\bar\sigma^{(n)}$,
\begin{subequations}
\label{eq:Delta_sB}
\begin{eqnarray}
\int_{-1}^1 dx\, x^{n-1} \Delta s_\Lambda(x)
&=& \frac12 \big( \bar\alpha^{(n)} + 2 \bar\sigma^{(n)} \big), \\
\int_{-1}^1 dx\, x^{n-1} \Delta s_{\Sigma^+}(x)
&=& \frac16 \big( \bar\alpha^{(n)} + 4 \bar\beta^{(n)} + 6 \bar\sigma^{(n)} \big) \\
&=& \int_{-1}^1 dx\, x^{n-1} \Delta s_{\Sigma^0}(x).
\end{eqnarray}
\end{subequations}
For the kaon tadpole distributions $\Delta s_K^{\rm (tad)}$ [Fig.~\ref{fig:loop-octet-decuplet}(b)], 
the moments are given by
\begin{subequations}
\label{eq:Delta_s_tad}
\begin{eqnarray}
\int_{-1}^1 dx\, x^{n-1} \Delta s_{K^+}^{\rm (tad)}(x)
&=& \frac{1}{12} \big( 5 \bar\alpha^{(n)} + 2 \bar\beta^{(n)} \big),	\\
\int_{-1}^1 dx\, x^{n-1} \Delta s_{K^0}^{\rm (tad)}(x)
&=& \frac{1}{12} \big( \bar{\alpha}^{(n)} + 4 \bar\beta^{(n)} \big).
\end{eqnarray}
\end{subequations}
For the distributions associated with the Kroll-Ruderman diagram [Fig.~\ref{fig:loop-octet-decuplet}(c)], the presence of the additional pion at the interaction vertex means that the moments of $\Delta s^{(\mathrm{KR})}_B$ are given in terms of the coefficients $\alpha^{(n)}$, $\beta^{(n)}$ and (in principle) $\sigma^{(n)}$,
\begin{subequations}
\label{eq:Delta_s_KR}
\begin{eqnarray}
\int_{-1}^1 dx\, x^{n-1} \Delta s^{(\mathrm{KR})}_\Lambda(x)
&=& \frac{\sqrt3}{4}\, \alpha^{(n)},	\\
\int_{-1}^1 dx\, x^{n-1} \Delta s^{(\mathrm{KR})}_{\Sigma^+}(x)
&=& \frac{1}{6\sqrt2} \big( \alpha^{(n)} + 4 \beta^{(n)} \big)      \\
&=& \sqrt{2} \int_{-1}^1 dx\, x^{n-1} \Delta s^{(\mathrm{KR})}_{\Sigma^0}(x).
\end{eqnarray}
\end{subequations}
Using SU(3) flavor symmetry, the axial vector and vector coefficients can also be written in terms of the spin-dependent and spin-averaged PDFs in the proton~\cite{prd-2016},
\begin{subequations}
\label{eq:alpha-beta-sigma-bar}
\begin{eqnarray}
\bar\alpha^{(n)}
&=& \frac13 \int_{-1}^1 dx\, x^{n-1}\, \big( 4 \Delta u(x) - 2 \Delta d(x) \big), \\
\bar\beta^{(n)}
&=& \frac13 \int_{-1}^1 dx\, x^{n-1}\, \big( 5 \Delta d(x) - \Delta u(x) \big), \\
\bar\sigma^{(n)} &=& 0,
\end{eqnarray}
\end{subequations}
and
\begin{subequations}
\label{eq:alpha-beta-sigma}
\begin{eqnarray}
\alpha^{(n)}
&=& \frac13 \int_{-1}^1 dx\, x^{n-1}\, \big( 4 u(x) - 2 d(x) \big) , \\
\beta^{(n)}
&=& \frac13 \int_{-1}^1 dx\, x^{n-1}\, \big( 5 d(x)  - u(x) \big), \\
\sigma^{(n)} &=& 0,
\end{eqnarray}
\end{subequations}
respectively.
From the relations in Eqs.~(\ref{eq:Delta_sB})--(\ref{eq:alpha-beta-sigma}) one can then write the spin-dependent strange quark PDFs $\Delta s_B$ and $\Delta s_K^{(\rm tad)}$ in the strange octet baryons in terms of the polarized nonstrange PDFs in the proton,
\begin{subequations}
\label{eq:slambda-ssigma}
\begin{eqnarray}
\Delta s_\Lambda(x)
&=& \frac13\, \big( 2 \Delta u(x) - \Delta d(x) \big), \\
\Delta s_{\Sigma^+}(x)
&=& \Delta s_{\Sigma^0}(x) = \Delta d(x),
\end{eqnarray}
\end{subequations}
and
\begin{subequations}
\label{eq:s-tad}
\begin{eqnarray}
\Delta s_{K^+}^{\rm (tad)}(x) &=& \frac12 \Delta u(x),	\\
\Delta s_{K^0}^{\rm (tad)}(x) &=& \frac12 \Delta d(x),
\end{eqnarray}
\end{subequations}%
and the spin-dependent strange Kroll-Ruderman PDFs $\Delta s^{(\mathrm{KR})}_B$ in terms of the unpolarized nonstrange PDFs in the proton,
\begin{subequations}
\label{eq:s-KR}
\begin{eqnarray}
\Delta s^{(\mathrm{KR})}_\Lambda(x)
&=& \frac{1}{2\sqrt3}\, \big( 2 u(x) -  d(x) \big),	\\
\Delta s^{(\mathrm{KR})}_{\Sigma^+}(x)
&=& \sqrt2\, \Delta s^{(\mathrm{KR})}_{\Sigma^0}(x) = \frac{1}{\sqrt2}\, d(x).
\end{eqnarray}
\end{subequations}

For the PDFs involving decuplet baryons, the moments of the spin-dependent distributions $\Delta s_T$ [Fig.~\ref{fig:loop-octet-decuplet}(d)] are related to the coefficient $\bar\gamma^{(n)}$ in Eq.~(\ref{eq:Os}),
\begin{eqnarray}
\label{eq:Delta_s_T}
\int_{-1}^1 dx\, x^{n-1} \Delta s_{\Sigma^{*+}}(x)
&=&  -  \frac13 \bar\gamma^{(n)}
 =  \int_{-1}^1 dx\, x^{n-1} \Delta s_{\Sigma^{*0}}(x),
\end{eqnarray}
while for the octet-decuplet transitions [Fig.~\ref{fig:loop-octet-decuplet}(e)] the moments of $\Delta s_{TB}$ are expressed in terms of the coefficient $\bar\omega^{(n)}$,
\begin{equation}
\label{eq:Delta_s_TB}
\int_{-1}^1 dx\, x^{n-1} \Delta s_{\Sigma^{*+}\Sigma^+}(x)
= - \frac{1}{\sqrt3}\, \bar\omega^{(n)}
= - \int_{-1}^1 dx\, x^{n-1} \Delta s_{\Sigma^{*0}\Sigma^0}(x).
\end{equation}
From SU(6) symmetry the coefficient $\bar\gamma^{(1)}$ can be related to the meson-baryon coupling constant $D$~\cite{Shanahan:2013},
\begin{equation}
\bar\gamma^{(1)} = - 3 D,
\end{equation}
from which the decuplet spin-dependent strange PDFs can be expressed as
\begin{equation}
\label{eq:sSig*}
\Delta s_{\Sigma^{*+}}(x)
= \Delta s_{\Sigma^{*0}}(x)
=  \frac12 \big( \Delta u(x) - 2 \Delta d(x) \big).
\end{equation}
For the coefficient of the octet-decuplet transition operators in Eq.~(\ref{eq:Os}), SU(3) symmetry gives the relation
\begin{eqnarray}
\bar{\omega}^{(n)}
&=& - \frac12\, \bar\alpha^{(n)} + \bar\beta^{(n)},
\end{eqnarray}
which allows the spin-dependent strange transition PDFs to be written as
\begin{equation}
\label{eq:sSig*Sig}
\Delta s_{\Sigma^{*+} \Sigma^+}(x)
= - \Delta s_{\Sigma^{*0} \Sigma^0}(x)
= \frac{1}{\sqrt3} \big( \Delta u(x) - 2 \Delta d(x) \big).
\end{equation}

With these relations, we have expressed all of the necessary strange quark distributions in the hadronic configurations in Fig.~\ref{fig:loop-octet-decuplet} in terms of PDFs in the bare proton, which, together with the hadronic splitting functions, constitute the input to the convolution formula in Eq.~(\ref{eq:convolution-1}).
In the next section we will derive the complete set of the hadronic splitting functions necessary to complete the evaluation of the PDFs.

\section{Hadronic splitting functions}
\label{sec:fy}

The spin-dependent hadronic splitting functions $\Delta f_j$ defined in Eq.~(\ref{eq:Delta_fj}) can be evaluated from the matrix elements of the hadronic operators in Eqs.~(\ref{eq:hadronic-operators})--(\ref{eq:hadronic-operators-T}), which correspond to the one meson loop diagrams in Fig.~\ref{fig:loop-octet-decuplet}.
In this section we derive each of the splitting functions for the octet rainbow, tadpole, octet Kroll-Ruderman, decuplet rainbow, and octet-decuplet transition contributions as a function of the light-cone variable $y=k^+/p^+$, where $k^\mu$ is the four-momentum of the kaon and $p^\mu$ is the four-momentum of the external proton.
The octet rainbow splitting functions have previously been computed in the literature \cite{Holtmann96, Malheiro97}, while the spin-dependent splitting functions for the tadpole and Kroll-Ruderman diagrams are computed here for the first time.

\subsection{Octet baryon rainbow} 
\label{ssec:fyH}

For the meson--octet baryon rainbow diagram of Fig.~\ref{fig:loop-octet-decuplet}(a), the splitting function is given by
\begin{eqnarray}\label{eq:fH-0}
\Delta f_{B\phi}^{(\rm rbw)}(y)
&=& \frac{1}{2M s^+} \frac{C^2_{B\phi}}{f_\phi^2}
    \int\!\frac{d^4 k}{(2\pi)^4}\,
    \bar{u}(p) (\slashed{k} \gamma_5)
    \frac{i(\slashed{p} - \slashed{k} + M_B)}{D_B}
    \gamma^+ \gamma_5
    \frac{i(\slashed{p} - \slashed{k} + M_B)}{D_B}
    (\gamma_5 \slashed{k})\, u(p)			\nonumber\\
&& \hspace*{3.6cm} \times\ \frac{i}{D_\phi} \delta(k^+ - y p^+),
\end{eqnarray}
where $D_\phi$ and $D_B$ are the meson and octet baryon virtualities,
\begin{subequations}
\begin{eqnarray}
D_\phi &=& k^2 - m_\phi^2 + i\epsilon,		\\
D_B &=& (p-k)^2 - M_B^2 + i\epsilon,
\end{eqnarray}
\end{subequations}
with $m_\phi$ and $M_B$ the kaon and octet baryon masses, respectively.
The spinor $u(p)$ is normalized such that $\bar{u}(p)\, u(p) = 2 M$, and $s^+$ is the ``+'' component of the external proton spin vector $s^\mu$.
The coefficients $C_{B\phi}^2$ can be obtained from the effective Lagrangian (\ref{eq:chiral-lagrangian}), and for the $\Lambda K$ and $\Sigma K$ configurations are explicitly given in terms of the $D$ and $F$ couplings as
\begin{equation}
\label{eq:C-B-phi}
C_{\Lambda K^+} = \frac{D+3F}{2\sqrt3},\ \ \ \
C_{\Sigma^+ K^0} = \sqrt2\, C_{\Sigma^0 K^+} = \frac{F-D}{\sqrt2}.
\end{equation}
Using the Dirac equation, the integrand in Eq.~(\ref{eq:fH-0}) can be decomposed into several terms with different combinations of meson and octet baryon propagators,
\begin{eqnarray}
\label{eq:fH-1}
\Delta f_{B\phi}^{(\rm rbw)}(y)
&=& - \frac{i}{2M s^+} \frac{C^2_{B\phi}}{f_\phi^2}
    \int\!\frac{d^4 k}{(2\pi)^4}\,
    \bigg[ \frac{N_1^B}{D^2_B D_\phi} 
         + \frac{N_2^B}{D_B D_\phi} 
         + \frac{N_3^B}{D_\phi}
    \bigg]\, \delta\Big( y - \frac{k^+}{p^+} \Big),
\end{eqnarray}
where
\begin{subequations}
\label{eq:ABC-octet-rbw}
\begin{eqnarray}
N_1^B
&=& - 2 \overline{M}_{\!B}^2 
    \Big [M \Delta_B^2\, s^+
        + 2 \Delta_B\, (k\!\cdot\!p\, s^+ - k\!\cdot\!s\, p^+)
        + M (k^2\, s^+ - 2\, k\!\cdot\!s\, k^+)
    \Big],  \\
N_2^B
&=& - 4 \overline{M}_{\!B} 
    \Big[ M \Delta_B\, s^+ 
        + (k\!\cdot\!p\, s^+ - k\!\cdot\!s\, p^+) 
    \Big], \\
N_3^B &=& - 2 M s^+,
\end{eqnarray}
\end{subequations}
with
\begin{eqnarray}
\label{eq:Bmassdefs}
\Delta_B \equiv M_B - M,\ \ \ \ \ \ \overline{M}_{\!B} \equiv M_B + M.
\end{eqnarray}
In a frame of reference in which $p_\perp = 0$, the two combinations
    $(k\!\cdot\!p\, s^+ - k\!\cdot\!s\, p^+)$ and
    $(k^2\, s^+ - 2\, k\!\cdot\!s\, k^+)$
appearing in Eqs.~(\ref{eq:ABC-octet-rbw}) become independent of $k^-$.
After integration over $k^+$, these two terms take the forms $y M^2 s^+$ and $(y^2 M^2 - k^2_\perp)\, s^+$, respectively.
It is convenient, therefore, to write the total octet baryon rainbow function $\Delta f_{B\phi}^{(\rm rbw)}$ as a sum of three splitting functions associated with the on-shell, off-shell and $\delta$-function contributions,
\begin{equation}
\label{eq:Delta_BK_rbw}
\Delta f_{B\phi}^{(\rm rbw)}(y)
= \frac{C_{B\phi}^2 \overline{M}_{\!B}^2}{(4\pi f_\phi)^2}
  \left[ \Delta f_B^{\rm (on)}(y) + \Delta f_B^{(\rm off)}(y) 
  + \Delta f_B^{(\delta)}(y)
  \right].
\end{equation}
Integrating over the $k^-$ component in Eq.~(\ref{eq:fH-1}) and using the residue theorem, one can write the individual functions in Eq.~(\ref{eq:Delta_BK_rbw}) in terms of integrals over $k_\perp^2$.
In particular, for the on-shell function one has
\begin{eqnarray}\label{eq:fH-on}
\Delta f^{(\mathrm{on})}_B(y) 
&=& y \int\!d k^2_\perp 
    \frac{\big[ - k^2_\perp + (\Delta_B + y M)^2 \big]}{\yb^2 D^2_{B\phi}} F^{(\mathrm{on})}_B(y,k^2_\perp),
\end{eqnarray}
where 
\begin{equation}
\label{eq:D_Bphi}
D_{B\phi} = - \frac{k^2_\bot + y M_B^2 + \yb\, m_\phi^2 - y \yb\, M^2}{\yb},
\end{equation}
and $F^{(\mathrm{on})}_B(y,k_\perp^2)$ is a function that represents the regularization of the $k_\perp^2$ integration (see Sec.~\ref{sec:regularization} below).

The result in Eq.~(\ref{eq:fH-on}) for the on-shell splitting function is in agreement with that in Refs.~\cite{Holtmann96, Malheiro97}.
On the other hand, the new off-shell splitting function in Eq.~(\ref{eq:Delta_BK_rbw}) is given by
\begin{equation}
\label{eq:Deltaf_B-off}
\Delta f_B^{(\mathrm{off})}(y)
= \frac{2}{\overline{M}_{\!B}\!} \int\!dk_\perp^2
  \frac{\big( \Delta_B + y M \big)}{\yb D_{B\phi}} F^{(\mathrm{off})}_B(y,k^2_\perp),
\end{equation}
where here $F^{(\mathrm{off})}_B(y,k_\perp^2)$ is the corresponding regulating function for the $k_\perp^2$ integration (which can in practice be different from the on-shell regulating function $F^{(\mathrm{on})}_B$ in Eq.~(\ref{eq:fH-on})).
For the $\delta$-function term, $\Delta f_\phi^{(\delta)}$, which arises from meson loops with zero light-cone momentum ($k^+ = 0$), one has
\begin{equation}
\Delta f_B^{(\delta)}(y)
= - \frac{1}{\overline{M}_{\!B}^2}\, \delta(y) \int\!dk_\perp^2\,
  \log\Omega_\phi\, F^{(\delta)}_B(y,k_\perp^2),
\label{eq:fKdel}
\end{equation}
where
    $\Omega_\phi = k_\bot^2 + m_\phi^2$, 
and $F^{(\delta)}_B(y,k_\perp^2)$ is the corresponding regulating function.

Compared with the splitting functions for the spin-averaged case derived in Ref.~\cite{prd-2016}, the spin-dependent on-shell function $\Delta f_B^{(\rm on)}$ in Eq.~(\ref{eq:fH-on}) differs from the spin-averaged analog by a change in sign of the $k_\perp^2$ term in the numerator of the integrand.
On the other hand, the off-shell function $\Delta f_B^{(\rm off)}$ and the $\delta$-function term $\Delta f_B^{(\delta)}$ are identical to the corresponding spin-averaged counterparts.

\subsection{Tadpole} 

The distribution functions associated with the meson tadpole diagram in Fig.~\ref{fig:loop-octet-decuplet}(b), with an operator insertion at the two nucleon--two meson vertex, can be written as
\begin{eqnarray}\label{eq:f-tad-K+}
\Delta f_\phi^{\rm (tad)}(y)
&=& \frac{1}{2 M s^+} \frac{1}{ f_{\phi}^2} \int\!\frac{d^4 k}{(2\pi)^4}\,
    \bar{u}(p) \gamma^+ \gamma_5\, u(p) 
    \frac{i}{D_\phi} \delta(k^+ - y p^+).
\end{eqnarray}
The tadpole splitting functions for the charged and neutral kaon loop contributions are then given by
\begin{equation}
\Delta f_{K^+}^{\rm (tad)}(y) = \Delta f_{K^0}^{\rm (tad)}(y)
\equiv - \frac{\overline{M}_{\!B}^2}{(4\pi f_{\phi})^2}\,
    \Delta f_\phi^{(\delta)}(y),
\end{equation}
where the generic tadpole function $\Delta f_\phi^{(\delta)}$ 
 related to the $\delta$-function term in the rainbow diagram in Eq.~(\ref{eq:fKdel}) is
\begin{equation}
\Delta f_\phi^{(\delta)}(y) = - \Delta f_B^{(\delta)}(y).
\end{equation}

\subsection{Kroll-Ruderman} 

The light-cone momentum distribution associated with the Kroll-Ruderman diagrams in Fig.~\ref{fig:loop-octet-decuplet}(c), which arise from the derivative coupling in the pseudovector chiral effective theory, is given by
\begin{eqnarray}\label{eq:f-KR}
\Delta f^{\rm (KR)}_{B\phi}(y)
&=& - \frac{i}{2M s^+} \frac{ C_{B\phi} }{f_\phi^2} \int\!\frac{d^4 k}{(2\pi)^4}\,
    \bar{u}(p)
    \left[ \slashed{k}\gamma_5
	   \frac{i(\slashed{p} - \slashed{k} + M_B)}{D_B}
	   \gamma^+ 
	 + \gamma^+
	   \frac{i(\slashed{p} - \slashed{k} + M_B)}{D_B}
	   \slashed{k} \gamma_5 \right]
    u(p)	\nonumber\\
&& \hspace*{3.9cm} \times\ \frac{i}{D_\phi} \delta( k^+ - y p^+).
\end{eqnarray}
Straightforward calculation gives
\begin{eqnarray}
\Delta f^{\rm (KR)}_{B\phi}(y)
&=& - \frac{i}{2 M s^+} \frac{C_{B\phi}}{f_{\phi}^2} \int\!\frac{d^4 k}{(2\pi)^4}\,
    \frac{4 \overline{M}_{\!B} (k\!\cdot\!p\, s^+ - k\!\cdot\!s\, p^+) 
          - 4 M (2 k\!\cdot\!p - k^2)\, s^+}
         {D_B D_\phi}                           \nonumber\\
& & \hspace{3.9cm} \times\, \delta(k^+ - y p^+).
\end{eqnarray}
The Kroll-Ruderman splitting function can then be written in terms of the off-shell and $\delta$-function contributions as
\begin{eqnarray}
\Delta f^{\rm (KR)}_{B\phi}(y)
&=&  - \frac{C_{B\phi}\, \overline{M}_{\!B}^2}{(4\pi f_\phi)^2}\,
 \Big[ \Delta f_B^{(\rm off)}(y) + 2 \Delta f_B^{(\delta)}(y) \Big],
\end{eqnarray}
with the off-shell function $\Delta f_B^{(\rm off)}$ as in Eq.~(\ref{eq:Deltaf_B-off}) and the $\delta$-function component $\Delta f_B^{(\delta)}$ in Eq.~(\ref{eq:fKdel}).

\subsection{Decuplet baryon rainbow}

For the decuplet intermediate states, because of the higher spin of the baryon the polarized splitting functions are somewhat more complicated.
The splitting function associated with the decuplet rainbow diagram in Fig.~\ref{fig:loop-octet-decuplet}(d) can be written as
\begin{eqnarray}
\label{eq:decuplet-rb}
\Delta f_{T\phi}^{(\rm rbw)}(y)
&=& \frac{1}{2Ms^+}
    \frac{C_{T\phi}^2}{f_\phi^2} \int\!\frac{d^4k}{(2\pi)^4}\, 
    \bar{u}(p)\, k_\mu \Theta^{\mu\rho}\, 
    \frac{i(\slashed{p}-\slashed{k}+M_T)}{D_T}
    {\cal P}^{\rho\alpha}(p-k)
    \gamma^+ \gamma_5 
    \frac{i(\slashed{p}-\slashed{k}+M_T)}{D_T} \nonumber\\
& & \hspace*{2.5cm} \times\
    {\cal P}^{\alpha\beta}(p-k)
    \Theta^{\beta\nu}\, k_\nu\, u(p) 
    \frac{i}{D_\phi} \delta(k^+ - y p^+),
\end{eqnarray}
where the usual spin-3/2 Rarita-Schwinger energy projector is
\begin{equation}
{\cal P}^{\alpha\beta}(p)
= g^{\alpha\beta} 
- \frac13 \gamma^\alpha \gamma^\beta
- \frac{1}{3 M_T}(\gamma^\alpha p^\beta - p^\alpha \gamma^\beta) 
- \frac{2}{3 M^2_T}\, p^\alpha p^\beta.
\end{equation}
This expression for the decuplet propagator corresponds to the particular choice $Z=1/2$ in Eq.~(\ref{eq:Theta}), for which the octet-decuplet transition tensor operator $\Theta^{\mu\nu}$ takes the simple form $g^{\mu\nu}-\gamma^\mu \gamma^\nu$.
The coefficients $C_{T\phi}^2$ can be derived from the effective Lagrangian~(\ref{eq:phi-B-T}), and for the $\Sigma^{*0} K$ and $\Sigma^{*+} K$ configurations are explicitly given by
\begin{equation}
\label{eq:C-T-phi}
C_{\Sigma^{*+} K^0} = -\sqrt2\, C_{\Sigma^{*0} K^+} = \frac{\cal C}{\sqrt6}.
\end{equation}
In our analysis, we will take ${\cal C} = - 2 D$ from SU(6) symmetry.
Straightforward but tedious calculation then allows $\Delta f_{T\phi}^{(\rm rbw)}$ to be written in a form similar to the octet baryon result in Eq.~(\ref{eq:fH-1}), 
\begin{eqnarray}
\label{eq:decup-rbw}
\Delta f_{T\phi}^{(\rm rbw)}(y)
&=& - \frac{i}{2M s^+} 
    \frac{C_{T\phi}^2}{f_\phi^2}
\int\!\frac{d^4 k}{(2\pi)^4}
\bigg[ \frac{N_1^T}{D^2_T D_\phi}
     + \frac{N_2^T}{D_T D_\phi}
     + \frac{N_3^T}{D_\phi}
\bigg]\, \delta\Big(y-\frac{k^+}{p^+}\Big),
\end{eqnarray}
as a sum of three terms involving different numbers of decuplet baryon propagator, $D_T$.
In analogy with the octet baryon splitting function in Eqs.~(\ref{eq:fH-1}) and (\ref{eq:ABC-octet-rbw}), the numerators $N_i^T$ in Eq.~(\ref{eq:decup-rbw}) can be written as linear combinations of the structures
  $2M s^+$,
  $(p\!\cdot\!k\, s^+ - k\!\cdot\!s\, p^+)$
and
  $(k^2\, s^+ - 2 k\!\cdot\!s\, k^+)$,
\begin{subequations}
\label{eq:NiT}
\begin{eqnarray}
N_1^T
&=& -\frac{2}{(3 M_T)^2}
    \Big[ \big(2 M M_T + \overline{M}_{\!T}^2\big) (p \cdot k)^2 
        + M \big(M^3_T - 4 M M^2_T - 7 M^2 M_T - 2 M^3\big)\, p \cdot k
    \nonumber\\
& & \hspace*{1.8cm}
    - M^2 \overline{M}_{\!T}^2\big(M_T + \overline{M}_{\!T}\big) \Delta_T
    \Big]\, 2 M s^+
    \nonumber\\
&+& \frac{8}{(3 M_T)^2}
    \Big[ \big(M_T + \overline{M}_{\!T}\big) p \cdot k\, 
          \big(p \cdot k - M \overline{M}_{\!T}\big)
        - M^2 \overline{M}_{\!T}^2 \Delta_T 
    \Big] 
    (p\!\cdot\!k\, s^+ - k\!\cdot\!s\, p^+)                   \nonumber\\
&-& \frac{4 M}{(3 M_T)^2}
    \Big[ (p \cdot k)^2 + M (\Delta_T - M)\, p \cdot k
        - M^2 \overline{M}_{\!T} (M_T + \Delta_T)
    \Big] (k^2\, s^+ -2 k\!\cdot\!s\, k^+),
    \nonumber\\
& & \\
N_2^T
&=& \frac{1}{(3 M_T^2)^2}
    \Big[ 4 \overline{M}_{\!T} \big(M_T+\overline{M}_{\!T}\big) (p \cdot k)^2
        - 2 M \big(4 M^3 + 12 M^2 M_T + 7 M M^2_T + M^3_T \big)\, p \cdot k
    \nonumber\\
& & \hspace*{1.4cm}
    + M^2 \big(4 M^4 + 12 M^3 M_T + 5 M^2 M^2_T - 6 M M^3_T - 3 M^4_T\big) \Big]\, 2 M s^+
    \nonumber\\
&-& \frac{4}{(3 M_T^2)^2}
    \Big[ 2\big(M_T+2\overline{M}_{\!T}\big) (p \cdot k)^2
        - M \big(8 M^2 + 12 M M_T - M^2_T\big)\, p \cdot k 
    \nonumber\\ 
& & \hspace*{1.4cm}
    + 2 M^2 \big(2 M^3 + 3 M^2 M_T - M M^2_T - 2 M^3_T\big)
    \Big] (p\!\cdot\!k\, s^+ - k\!\cdot\!s\, p^+)
    \nonumber\\ 
&+& \frac{2 M}{(3 M_T^2)^2}
    \Big[ 4 \big(M^2 - p \cdot k\big)^2 
        - M^2 M^2_T
    \Big] (k^2\, s^+ - 2 k\!\cdot\!s\, k^+),    \\
& & \nonumber\\
N_3^T
&=& \frac{1}{(3 M_T^2)^2}
    \Big[ \big(M_T+2\overline{M}_{\!T}\big)^2 k^2 
        - M^2 \big( 4 M^2 + 12 M M_T + 7 M^2_T\big)
    \Big] 2 M s^+
    \nonumber\\
&+& \frac{8}{(3 M_T^2)^2} 
    \Big[ \big(M_T+2\overline{M}_{\!T}\big) \big(M^2 - p\!\cdot\!k\big) \Big]
    (p\!\cdot\!k\, s^+ - k\!\cdot\!s\, p^+),
\end{eqnarray}
\end{subequations}
where we define the difference and sum of the masses for the decuplet baryons as in Eq.~(\ref{eq:Bmassdefs}),
\begin{eqnarray}
\Delta_T \equiv M_T - M,\ \ \ \ \ \ \ \
\overline{M}_{\!T} \equiv M_T + M.
\end{eqnarray}
This structure then allows the decuplet rainbow splitting function to be  decomposed into decuplet on-shell, off-shell and $\delta$-function terms,
\begin{equation}\label{eq:f-TK-rbw}
\Delta f_{T\phi}^{(\rm rbw)}(y)
= \frac{C_{T\phi}^2 \overline{M}_{\!T}^2}{(4\pi f_\phi)^2} 
\Big[ \Delta f^{(\rm on)}_T(y)
    + \Delta f^{(\rm off)}_T(y)
    + \Delta f_T^{(\delta)}(y)
\Big].
\end{equation}
Details of the derivations of the individual functions in Eq.~(\ref{eq:f-TK-rbw}) are given in Appendix~\ref{sec:fT}.
After the $k^-$ integration we therefore obtain
\begin{eqnarray}
\label{eq:fT-on}
\Delta f^{(\rm on)}_T(y)  
&=& - \frac{1}{2 \big( 3 M_T \overline{M}_{\!T} \big)^2} 
\int d k^2_\bot\, \dfrac{y}{\yb^4 D_{T\phi}^2}
\bigg\{ 
  \Big[ k_\bot^2 + \big(M_T + \yb M \big)^2 \Big]
\nonumber\\ 
&& \hspace*{2.2cm} \times 
  \Big[ k_\bot^4 - 8 \yb M M_T\, k_\bot^2 - \big( M_T^2 - \yb^2 M^2 \big)^2
  \Big] 
\bigg\}\,
F_T^{({\rm on})}(y, k_\bot^2),
\end{eqnarray}
and
\begin{eqnarray}
\label{eq:fT-off}
\Delta f^{(\rm off)}_T(y)
&=& \frac{1}{\big(3 M_T^2 \overline{M}_{\!T} \big)^2} 
\int d k^2_\bot\,
\dfrac{1}{\yb^3 D_{T\phi}} 
\bigg\{
  k_\bot^6
- \Big[ M_T^2 + 3 \yb M M_T - \yb^2 M^2 \Big] k_\bot^4 
\nonumber\\
&& \hspace*{1.8cm} 
- \Big[ 3 M_T^4 
      + 2 \yb M M_T^3 
      + 4 \yb^2 M^2 M_T^2 
      + 6 \yb^3 M^3 M_T 
      + \yb^4 M^4 
  \Big] k_\bot^2
\nonumber\\
&& \hspace*{1.8cm} 
- \Big[ M_T^3 - 2 \yb M M_T^2 + \yb^3 M^3 \Big] 
  \big( M_T + \yb M \big)^3
\bigg\}
F_T^{({\rm off})}(y, k_\bot^2),
\end{eqnarray}
for the decuplet on-shell and off-shell functions, respectively, with $F_T^{({\rm on})}$ and $F_T^{({\rm off})}$ the corresponding regulating functions, and in analogy with Eq.~(\ref{eq:D_Bphi}) we have
\begin{equation}
\label{eq:D_Tphi}
D_{T\phi} = - \frac{k^2_\bot + y M_T^2 + \yb\, m_\phi^2 - y \yb\, M^2}{\yb}.
\end{equation}
For the $\delta$-function contribution, we have
\begin{eqnarray}
\label{eq:fT-del}
\Delta f^{(\delta)}_T(y)
&=& \frac{1}{\big(3 M_T \overline{M}_{\!T}\big)^2}
\bigg\{ 
  \Big[ \big(M_T+2\overline{M}_{\!T}\big)^2 m_\phi^2 - M^2 \big(4 M^2 + 12 M M_T + 7 M_T^2\big) \Big]
  \Delta f_1^{(\delta)}(y)
\nonumber\\
&& \hspace*{1.6cm} 
  - \Big[ 2 M (M_T+2\overline{M}_{\!T}) \Big] \Delta f_2^{(\delta)}(y)
\bigg\},
\end{eqnarray}
where the two functions proportional to $\delta(y)$ are given by
\begin{subequations}
\begin{eqnarray}
\Delta f_1^{(\delta)}(y)
&=& \delta(y) \frac{1}{M_T^2} \int\!dk^2_\bot\,
\log\Omega_\phi\,
F_T^{(\delta 1)}(y,k_{\bot}^2),
\label{eq:Deltaf1factor} \\
\Delta f_2^{(\delta)}(y)
&=& \delta(y) \frac{1}{M_T^2} \int\!dk^2_\bot\,
\Omega_\phi \log\Omega_\phi\,
F_T^{(\delta 2)}(y,k_{\bot}^2),
\label{eq:Deltaf2factor}
\end{eqnarray}
\end{subequations}
with regulating functions $F_T^{(\delta 1)}(y,k_{\bot}^2)$ and $F_T^{(\delta 2)}(y,k_{\bot}^2)$, respectively.
Explicit expressions for each of the regulating functions are given in Sec.~\ref{sec:regularization} for Pauli-Villars regularization.

\subsection{Octet-decuplet baryon transition}

For the octet-decuplet rainbow transition diagrams in Fig.~\ref{fig:loop-octet-decuplet}(e), the splitting function can be written as
\begin{eqnarray}
\Delta f_{TB\phi}^{\rm (rbw)}(y)
&=& -\frac{1}{2M s^+}
\frac{C_{T\phi} C_{B\phi}}{f_\phi^2}
\int\!\frac{d^4 k}{(2\pi)^4}
\bar{u}(p)
\nonumber\\
&\times&
\bigg[
    k_\mu \Theta^{\mu\rho} 
    \frac{i (\slashed{p}-\slashed{k}+M_T)}{D_T}
    {\cal P}^{\rho\nu}(p-k)
    \Theta^{\nu +} 
    \frac{i (\slashed{p}-\slashed{k}+M_B)}{D_B}
    \slashed{k} \gamma_5
\nonumber\\
& & \hspace{-0.1cm}\
 +\, \slashed{k} \gamma_5 
    \frac{i (\slashed{p} - \slashed{k} + M_B)}{D_B}
    \Theta^{+ \mu} 
    \frac{i (\slashed{p} - \slashed{k} + M_T)}{D_T}
    {\cal P}^{\mu\alpha}(p-k) 
    \Theta^{\alpha\nu} k_\nu
    \bigg]
\nonumber\\
&\times& 
u(p) \frac{i}{D_\phi}
\delta\Big(y - \frac{k^+}{p^+}\Big),
\label{eq:fTBphi}
\end{eqnarray}
for the $TB\phi = \Sigma^{*0} \Sigma^0 K^+$ and $\Sigma^{^*+} \Sigma^+ K^0$ configurations, with $C_{B \phi} $ and $C_{T \phi}$ given by Eqs.~(\ref{eq:C-B-phi}) and~(\ref{eq:C-T-phi}), respectively.
The two terms in the brackets of Eq.~(\ref{eq:fTBphi}) correspond to the two 
orderings of $BT$ and $TB$ in Fig.~\ref{fig:loop-octet-decuplet}(e).
Also note that there are no Kroll-Ruderman type diagrams with decuplet intermediate states contributing to spin-dependent splitting functions.
In analogy with the splitting functions for the octet and decuplet baryon intermediate states in Eqs.~(\ref{eq:fH-1}) and (\ref{eq:decup-rbw}), we write the octet-baryon transition rainbow splitting function as a sum of three terms with different numbers of baryon propagators,
\begin{equation}
\label{eq:f-TB}
\Delta f_{TB\phi}^{\rm (rbw)}(y)
= \frac{i}{2M s^+} 
\frac{C_{T\phi} C_{B\phi}}{f_\phi^2}
\int\!\frac{d^4 k}{(2\pi)^4}
\bigg[\frac{N_1^{TB}}{D_T D_B D_\phi} 
    + \frac{N_2^{TB}}{D_B D_\phi} 
    + \frac{N_3^{TB}}{D_\phi} 
\bigg]
\delta\Big(y - \frac{k^+}{p^+}\Big),
\end{equation}
where the numerators of the terms in the brackets are given by
\begin{subequations}
\label{eq:NiTB}
\begin{eqnarray}
N_1^{TB}
&=& \frac{\overline{\!M}_T \overline{M}_{\!TB}}{3 M^2_T}
\bigg\{
\Big[ 2 \big(\overline{M}_{\!T}^2 - M M_T\big) p \cdot k
    + M \big(M + \overline{M}_{\!T}\big) \overline{M}_{\!T} \Delta_T 
\Big] 2 M s^+
\nonumber\\
& & \hspace*{1.7cm} 
- \Big[ 4 \big(M + \overline{M}_{\!T}\big) p \cdot k
      + 8 M \overline{M}_{\!T} \Delta_T
  \Big]
(p\!\cdot\!k\, s^+ - k\!\cdot\!s\, p^+)
\nonumber \\
& &  \hspace*{1.7cm} 
+ \Big[ 4M p \cdot k + 2M^2 \big( \Delta_T - M \big) \Big]
(k^2\, s^+ - 2  k\!\cdot\!s\, k^+)
\bigg\},                                     \\
&& \nonumber \\
N_2^{TB} 
&=& \frac{1}{3 M^2_T} 
\Big[ 2 \big( \Delta_{TB}^2 - M \overline{M}_{\!TB} - M^2 \big)\, p \cdot k 
    + 3 M_T \big( 2 M^2 M_B - M^3_B + M^2 M_T \big)
\nonumber\\
& & \hspace*{1cm}
+\, M \overline{M}_{\!TB} \big( 4 M^2 - 2 M^2_B + M^2_T \big) \Big]
2 M s^+
\nonumber\\
&+& \frac{4}{3 M^2_T}
\Big[ \big(M_B - \Delta_{TB}\big) \big(p \cdot k - M^2\big)
    + 2 M^3
    - 2 M \big(\overline{M}_{\!TB}^2 - M_B M_T\big)
    - 3 M^2_B M_T
\Big]
\nonumber\\
& & \hspace*{1cm}
\times\, (p\!\cdot\!k\, s^+ - k\!\cdot\!s\, p^+)
\nonumber\\
&+& \frac{2 M}{3 M^2_T} 
\Big[ 2\, p \cdot k - 2 M \overline{M}_{\!TB} - 3 M_B M_T - 4 M^2 \Big]
(k^2\, s^+ - 2 k\!\cdot\!s\, k^+),
\\
&& \nonumber\\
N_3^{TB}
&=& \frac{1}{3 M^2_T} 
\Big[ 2\, p \cdot k - 2 M \overline{M}_{\!TB} - 3 M_B M_T \Big]
2 M s^+
\nonumber\\ 
&-& \frac{4}{3 M^2_T}
\Big[ M_T + 2 \overline{M}_{\!T} \Big]
(p\!\cdot\!k\, s^+ - k\!\cdot\!s\, p^+),
\\
&& \nonumber
\end{eqnarray}
\end{subequations}
and we define
\begin{eqnarray}
\Delta_{TB} \equiv M_T - M_B,\ \ \ \ \ \ \overline{M}_{\!TB} \equiv M_T + M_B.
\end{eqnarray}
Finally, as with the octet-only and decuplet-only intermediate state contributions, the octet-decuplet transition splitting function can be written in terms of on-shell, off-shell and $\delta$-function terms,
\begin{equation}
\label{eq:f-TBK}
\Delta f_{TB\phi}^{\rm (rbw)}(y)
= \frac{C_{T\phi} C_{B\phi} \overline{M}_{\!T} \overline{M}_{\!TB}}{(4\pi f_\phi)^2}
\Big[
  \Delta f^{(\rm on)}_{TB}(y)
+ \Delta f^{(\rm off)}_{TB}(y)
+ \Delta f_{TB}^{(\delta)}(y)
\Big].
\end{equation}
Following the steps given in Appendix~\ref{sec:fT}, the on-shell octet-decuplet transition function in Eq.~(\ref{eq:f-TBK}) can be written as
\begin{eqnarray}
\label{eq:fTB-on}
\hspace*{-2cm}
\Delta f^{(\rm on)}_{TB}(y)
&=& \frac{1}{3 M_T^2 \overline{M}_{\!TB} \Delta_{TB}}
\int \frac{dk_\bot^2}{\yb^2}\,
\bigg( \frac{F_{TB}^{(T)}}{D_{T\phi}} 
     - \frac{F_{TB}^{(B)}}{D_{B\phi}}
\bigg)
\nonumber\\ 
&& \hspace*{3cm} \times
\Big[ 
  k_\bot^4
- \big( 2 M_T \Delta_{TB} + \yb M (3 M_T - M_B) \big) k_\bot^2
\nonumber\\ 
&& \hspace*{3.5cm}
- (\Delta_B + y M) (\Delta_T + y M )\big(\overline{M}_{\!T} - y M\big)^2
\Big],
\end{eqnarray}
where the regulator functions $F_{TB}^{(T)}$ and $F_{TB}^{(B)}$ are given in Sec.~\ref{sec:regularization} below.
The off-shell transition function is given by
\begin{eqnarray}
\label{eq:fTB-off}
\hspace*{-1cm}
\Delta f^{(\rm off)}_{TB}(y)
&=& \frac{1}{3 M_T^2 \overline{M}_{\!T} \overline{M}_{\!TB}} 
\int \frac{dk_\bot^2}{\yb^2}
\nonumber\\
&\times&
\bigg\{
\frac{F_{TB}^{(T)}}{D_{T\phi}}\,
\Big[\,
  \overline{M}_{\!T} \big(2 M_T + \yb M \big) k_\bot^2
- \overline{M}_{\!T} \big(\Delta_T + y M \big) 
    \big(\overline{M}_{\!T} - y M\big)^2
\Big]
\nonumber\\
& & \hspace*{0cm}
+ \frac{F_{TB}^{(B)}}{D_{B\phi}}\,
\Big[\,
k_\bot^4 
+ \big( M_T (\overline{M}_{\!B} - 2 \Delta_{TB})
     + \yb\, (3 M^2 + 4 M M_B + 3 M_B M_T 
  \big) k_\bot^2
\nonumber\\
& & \hspace*{1.3cm}
-\, (\Delta_B + y M) 
\big(
    M_B^3 
  + M_T^3 
  + (1+\yb) M \overline{M}_{\!T} \Delta_T
  + y \yb M^2 (4 M_T + M_B)
\nonumber\\
& & \hspace*{1.3cm}
  +\, M_B \overline{M}_{\!TB} (\overline{M}_{\!T} + \yb M) 
  - 3 y M_B M_T (M_B + y M)
\big)
\Big]
\bigg\}\, ,
\end{eqnarray}
in terms of the same regulators $F_{TB}^{(B)}$ and $F_{TB}^{(T)}$ as in the on-shell function (\ref{eq:fTB-on}).
Finally, for the $\delta$-function contribution to the octet-decuplet transition, we find 
\begin{eqnarray}
\label{eq:fTB-delta}
\Delta f^{(\delta)}_{TB}(y)
&=& \frac{1}{3 \overline{M}_{\!T}}
    \bigg( 2 M + \frac{3 M_B M_T}{\overline{M}_{\!TB}} \bigg)\,
    \Delta f_1^{(\delta)}(y),
\end{eqnarray}
where the function $\Delta f_1^{(\delta)}$ is given in Eq.~(\ref{eq:Deltaf1factor}).

\newpage
\section{Nonanalytic behavior}
\label{sec:LNA}

In the chiral expansion of moments of PDFs, the coefficients of the LNA terms in the pseudoscalar meson mass, $m_\phi$, are model independent and can only arise from meson loops.
Within the convolution framework of Sec.~\ref{ssec:convolution}, the LNA behavior of the nucleon PDF moments is determined by the LNA behavior of the moments of the splitting functions describing the transitions to the meson-baryon intermediate states.
In the unpolarized case, the LNA terms were previously found to have a characteristic $m_\phi^2 \log m_\phi^2$ dependence~\cite{Ji:2001, Thomas:2000, Detmold:2001jb, Savage:2002}.

To begin with, we define the $n{\rm th}$ moment of the spin-dependent splitting function $\Delta {\widetilde f}_{h\, (i)}^{\, (n)}$ in the hadronic configuration $h=B, T$ or $TB$ by
\begin{eqnarray}
\Delta {\widetilde f}_{h\, (i)}^{\, (n)}
&=& \int_0^1 dy\, y^{n-1}\, \Delta f_h^{(i)}(y),
\end{eqnarray}
for the $i=\{$on, off, $\delta \}$ contribution.
From the convolution expression for the $\Delta s$ PDF in the nucleon in Eqs.~(\ref{eq:convolution-1}) and (\ref{eq:convolution-2}), and the definition of the nucleon PDF moment in Eq.~(\ref{eq:mom_def}), we can write the $n{\rm th}$ moment of the strange PDF in the nucleon as
\begin{eqnarray}
\label{eq:binom}
\langle x^{n-1} \rangle_{\Delta s}
&=& 
\sum_{h, i}
\sum_{k=1}^n \binom{n\!-\!1}{k\!-\!1} (-1)^{k-1}\,
    \Delta {\widetilde f}_{h\, (i)}^{\, (k)}\,
    \Delta S_h^{(n-1)},
\end{eqnarray}
where
\begin{eqnarray}
\Delta S_h^{(n-1)} &=& \int_0^1 dx\, x^{n-1} \Delta s_h(x)
\end{eqnarray}
is the $n{\rm th}$ moment of the strange quark PDF $\Delta s_h$ in the hadronic configuration $h$.
The binomial symbol in Eq.~(\ref{eq:binom}) arises from the splitting functions in Eq.~(\ref{eq:convolution-2}) being evaluated at~$\yb$.
From the relations in Sec.~\ref{ssec:hadronic-pdfs}, the moments
    $\Delta S_h^{(n-1)}$ 
are given in terms of the coefficients $\bar\alpha^{(n)}$, $\bar\beta^{(n)}$, $\bar\sigma^{(n)}$, $\bar\gamma^{(n)}$, $\bar\omega^{(n)}$, $\alpha^{(n)}$ and $\beta^{(n)}$.
Writing the contributions from the different types of splitting functions in Fig.~\ref{fig:loop-octet-decuplet} explicitly, we can compute the LNA behavior of the strange PDF moments as
\begin{eqnarray}
\langle x^{n-1} \rangle_{\Delta s}^{\rm LNA}
&=& \sum_{B\phi}
\frac{\overline{M}_{\!B}^2}{(4\pi f_\phi)^2}
\sum_{k=1}^n \binom{n\!-\!1}{k\!-\!1} (-1)^{k-1}
\nonumber\\
& & \hspace*{0.5cm} \times
\bigg\{
  C_{B\phi}^2\,
  \Big[
    \Delta{\widetilde f}_{B ({\rm on})}^{\, (k)}
  + \Delta{\widetilde f}_{B ({\rm off})}^{\, (k)}
  + \Delta{\widetilde f}_{B (\delta)}^{\, (k)}
  \Big]_{\rm LNA}
  \Delta S^{(n-1)}_B
\nonumber\\
& &
\hspace*{0.7cm}
-\ C_{B\phi}\,
  \Big[
       \Delta{\widetilde f}_{B ({\rm off})}^{\, (k)}
   + 2 \Delta{\widetilde f}_{B (\delta)}^{\, (k)}
  \Big]_{\rm LNA}
  \Delta S_{B (\rm KR)}^{(n-1)}
\nonumber\\
& &
\hspace*{1.53cm}
+\ \Big[
        \Delta{\widetilde f}_{B (\delta)}^{\, (k)}
  \Big]_{\rm LNA}
  \Delta S_{\phi (\rm tad)}^{(n-1)}
\bigg\}
\nonumber\\
&+& \sum_{T\phi}
\frac{\overline{M}_{\!T}^2}{(4\pi f_\phi)^2}
\sum_{k=1}^n \binom{n\!-\!1}{k\!-\!1} (-1)^{k-1}
\nonumber\\
& & \hspace*{0.5cm} \times
\bigg\{
  C_{T\phi}^2\,
  \Big[
    \Delta{\widetilde f}_{T ({\rm on})}^{\, (k)}
  + \Delta{\widetilde f}_{T ({\rm off})}^{\, (k)}
  + \Delta{\widetilde f}_{T (\delta)}^{\, (k)}
  \Big]_{\rm LNA}
  \Delta S^{(n-1)}_T
\bigg\}
\nonumber\\
&+& \sum_{BT}
\frac{\overline{M}_{\!T} \overline{M}_{\!TB}}{(4\pi f_\phi)^2}
\sum_{k=1}^n \binom{n\!-\!1}{k\!-\!1} (-1)^{k-1}
\nonumber\\
& & \hspace*{0.5cm} \times
\bigg\{
  C_{B\phi} C_{T\phi}\,
  \Big[
    \Delta{\widetilde f}_{TB ({\rm on})}^{\, (k)}
  + \Delta{\widetilde f}_{TB ({\rm off})}^{\, (k)}
  + \Delta{\widetilde f}_{TB (\delta)}^{\, (k)}
  \Big]_{\rm LNA}
  \Delta S^{(n-1)}_{TB}
\bigg\}\, .
\label{eq:sLNA}
\end{eqnarray}

In the following we focus specifically on the $n=1$ moment of the strange quark PDF, $\langle x^0 \rangle_{\Delta s}^{\rm LNA} \equiv \Delta S^{(0)}_{\rm LNA}$, which requires computing the LNA behavior of the $n=1$ moments of the splitting functions, $\Delta {\widetilde f}_{h\, (i)}^{\, (1)}$.
These are expanded in powers of $m_\phi/M$, $\Delta_B/M$, and $\Delta_T/M$, and consider the nonanalytic (NA) behavior, which includes LNA and also higher powers, of the individual on-shell, off-shell and $\delta$-function contributions.
For the octet baryons, the NA behavior of the $n=1$ moment of the on-shell function is given for the cases when $\Delta_B > m_\phi$ or $\Delta_B < m_\phi$,
\begin{eqnarray}
\overline{M}_{\!B}^2\, \Delta{\widetilde f}_{B ({\rm on})}^{\, (1)}\Big|_{\rm NA}
= \left\{
\begin{array}{l}
  2 \Delta_B^2 \log m_{\phi}^2
- 2 R_B \Delta_B \log\dfrac{\Delta_B - R_B}{\Delta_B + R_B}\, ,
\hspace*{1.8cm} [\Delta_B > m_\phi]  \\
                                    \\
  2 \Delta_B^2 \log m_{\phi}^2
- 2 \overline{R}_B\, \Delta_B \Big( \pi - 2 \arctan\dfrac{\Delta_B}{\overline{R}_B} \Big)\, ,
\hspace*{0.7cm} [\Delta_B < m_\phi]
\end{array}
\right.
\end{eqnarray}
where
    $R_B = \sqrt{\Delta^2_B - m^2_\phi}$
and
    $\overline{R}_B = \sqrt{ m_{\phi}^2 - \Delta^2_B}$.
The spin-dependent off-shell and $\delta$-function terms are equivalent to the corresponding unpolarized splitting functions, and for the $n=1$ moments have the NA behavior~\cite{Salamu:2019-1},
\begin{eqnarray}
\overline{M}_{\!B}^2\, \Delta{\widetilde f}_{B ({\rm off})}^{\, (1)}\Big|_{\rm NA}
&=&
\left\{
 \begin{array}{l}
 - 2 m_\phi^2 \log m_\phi^2
 - \dfrac{2 R_B^3}{M_B} \log\dfrac{\Delta_B - R_B}{\Delta_B + R_B}\, ,
   \hspace*{2.1cm} [\Delta_B > m_\phi]  \\
                                    \\
 - 2 m_\phi^2 \log m_\phi^2
 + \dfrac{2 \overline{R}_B^3}{M_B}
   \Big( \pi - 2 \arctan\dfrac{\Delta_B}{\overline{R}_B}\Big)\, ,
\hspace*{1.1cm} [\Delta_B < m_\phi]
\end{array}
\right.    \\
& & \nonumber\\
\overline{M}_{\!B}^2\, \Delta{\widetilde f}_{B (\delta)}^{\, (1)}\Big|_{\rm LNA}
&=& m_\phi^2 \log m_\phi^2,
\label{eq:f1delLNA}
\end{eqnarray}
respectively.

For the decuplet rainbow splitting functions, the NA behavior of the $n=1$ moments of the on-shell and off-shell functions is given by
\begin{eqnarray}
\label{eq:TonLNA}
\overline{M}_{\!T}^2\, 
\Delta{\widetilde f}_{T ({\rm on})}^{\, (1)}\Big|_{\rm NA} 
&=&
\left\{
\begin{array}{l}
  \dfrac{4}{3} 
  \Big[ m_\phi^2 - \dfrac{\Delta_T}{6} \big( M + \frac{21}{2} \Delta_T \big) \Big] 
  \log m_\phi^2 
+ \dfrac{2 R_T}{9}
  \big( M + \frac{21}{2} \Delta_T \big) 
  \log\dfrac{\Delta_T - R_T}{\Delta_T + R_T},
\\
\hspace*{10.7cm} [\Delta_T > m_\phi]
\\
\\
  \dfrac{4}{3}
  \Big[ m_\phi^2 - \dfrac{\Delta_T}{6} \big( M  + \frac{21}{2} \Delta_T\big) \Big] 
  \log m_\phi^2
+ \dfrac{2 \overline{R}_T}{9}
  \big( M + \frac{21}{2} \Delta_T \big)\,
  \Big( \pi - 2 \arctan\dfrac{\Delta_T}{\overline{R}_T} \Big),
\\
\hspace*{10.7cm} [\Delta_T < m_\phi]
\end{array}
\right.
\end{eqnarray}
and
\begin{eqnarray}
\label{eq:ToffLNA}
\overline{M}_{\!T}^2\,
\Delta{\widetilde f}_{T ({\rm off})}^{\, (1)}\Big|_{\rm NA}
&=&
\left\{
\begin{array}{l}
- \dfrac{4}{9}
  \Big[ m_\phi^2 - \dfrac{\Delta_T}{2} \big( M + \frac{1}{2} \Delta_T \big) \Big]
  \log m_\phi^2 
- \dfrac{2 R_T}{9}
  \big( M + \frac{1}{2} \Delta_T \big)
  \log\dfrac{\Delta_T - R_T}{\Delta_T + R_T},
\\
\hspace*{10.7cm} [\Delta_T > m_\phi]
\\
\\
- \dfrac{4}{9}
  \Big[ m_\phi^2 - \dfrac{\Delta_T}{2} \big( M + \frac{1}{2} \Delta_T \big) \Big]
  \log m_\phi^2 
- \dfrac{2 \overline{R}_T}{9}
  \big( M + \frac{1}{2} \Delta_T \big)
  \Big( \pi - 2 \arctan\dfrac{\Delta_T}{\overline{R}_T} \Big),
\\
\hspace*{10.7cm} [\Delta_T < m_\phi]
\end{array}
\right.
\end{eqnarray}
respectively, where
    $R_T = \sqrt{\Delta^2_T - m^2_\phi}$
and
    $\overline{R}_T = \sqrt{ m_\phi^2 - \Delta^2_T}$.
Note that while the results for the individual on-shell and off-shell contributions in Eqs.~(\ref{eq:TonLNA}) and (\ref{eq:ToffLNA}) depend on the choice of the decomposition into the two pieces, 
the sum of the on-shell and off-shell contributions is independent of the separation, and gives rise to
\begin{eqnarray}
\label{eq:NA-fT-sum}
 \overline{M}_{\!T}^2\,
 \Big( \Delta{\widetilde f}_{T ({\rm on})}^{\, (1)}
     + \Delta{\widetilde f}_{T ({\rm off})}^{\, (1)}
 \Big)_{\rm NA}
&=&\left\{
\begin{array}{l}
  \dfrac89
  \Big[ m_\phi^2 - \dfrac{5}{2} \Delta_T^2 \Big] \log m_\phi^2 
+ \dfrac{20 \Delta_T R_T}{9}
  \log\dfrac{\Delta_T - R_T}{\Delta_T + R_T},
\\
\hspace*{8cm} [\Delta_T > m_\phi]
\\
\\
  \dfrac89
  \Big[ m_\phi^2 - \dfrac{5}{2} \Delta_T^2 \Big] \log m_\phi^2 
+ \dfrac{20 \Delta_T \overline{R}_T}{9}
  \Big( \pi - 2 \arctan\dfrac{\Delta_T}{\overline{R}_T} \Big).
\\
\hspace*{8cm} [\Delta_T < m_\phi]
\end{array}
\right.
\end{eqnarray}
The LNA contribution arising from the $\delta$-function term is given by
\begin{equation}
\overline{M}_{\!T}^2\,
\Delta{\widetilde f}_{T}^{(\delta)} \Big|_{\rm LNA}
= \frac{23}{9}\, m_\phi^2 \log m_\phi^2.
\end{equation}

For the octet-decuplet transition splitting functions, the NA behavior is slightly more involved because of the presence of two baryon mass differences, $\Delta_B$ and $\Delta_T$.
For the on-shell and off-shell splitting functions, the first moments are given by
\begin{eqnarray}
\overline{M}_{\!T} \overline{M}_{\!TB}\,
\Delta{\widetilde f}_{TB ({\rm on})}^{\, (1)}\Big|_{\rm NA}
&=&
\Big[ 
- 4 m_\phi^2 + 2M \big( \tfrac13 \Delta_B + \Delta_T \big)
+ \tfrac{25}{9} \Delta_B^2 + \tfrac{13}{9} \Delta_B \Delta_T + \tfrac{34}{9} \Delta_T^2
\Big]
\log m_\phi^2
\nonumber\\
& & -\
\overline{R}_B
\bigg( M_B - \tfrac13 M_T + \frac{16\, \overline{R}_B^2}{9\, \Delta_{TB}} \bigg)
\Big( \pi - 2 \arctan\dfrac{\Delta_B}{\overline{R}_B} \Big)
\nonumber\\
& & -\
\overline{R}_T
\bigg( 2 M_T - \frac{16\, \overline{R}_T^2}{9\, \Delta_{TB}} \bigg)
\Big( \pi - 2 \arctan\dfrac{\Delta_T}{\overline{R}_T} \Big),
\\
\overline{M}_{\!T} \overline{M}_{\!TB}\,
\Delta{\widetilde f}_{TB ({\rm off})}^{\, (1)}\Big|_{\rm NA}
&=&
\Big[
  \tfrac{8}{3} m_\phi^2 - 2M \big( \tfrac13 \Delta_B + \Delta_T \big)
- \Delta_B^2 + \tfrac{1}{3} \Delta_B \Delta_T - 2 \Delta_T^2 
\Big] 
\log m_\phi^2
\nonumber\\
& & +\
\overline{R}_B
\big( M_B - \tfrac13 M_T \big)
\Big( \pi - 2 \arctan\dfrac{\Delta_B}{\overline{R}_B} \Big)
\nonumber\\
& & +\
2 \overline{R}_T M_T 
\Big( \pi - 2 \arctan\dfrac{\Delta_T}{\overline{R}_T} \Big),
\end{eqnarray}
for $\Delta_B < m_\phi$ and $\Delta_T < m_\phi$.
There is strong cancellation between the on-shell and off-shell pieces, resulting in a sum given by
\begin{eqnarray}
\overline{M}_{\!T} \overline{M}_{\!TB}\,
\Big(
  \Delta{\widetilde f}_{TB ({\rm on})}^{\, (1)}
+ \Delta{\widetilde f}_{TB ({\rm off})}^{\, (1)}
\Big)_{\rm NA}
&=& 
\Big[
- \tfrac{4}{3} m_\phi^2
+ \tfrac{16}{9} \big( \Delta_B^2 + \Delta_B \Delta_T + \Delta_T^2 \big)
\Big]
\log m_\phi^2
\nonumber\\
&& \hspace*{-1cm}
-\, \dfrac{16}{9 \Delta_{TB}}
\bigg[
  \overline{R}_B^3 
  \Big( \pi - 2 \arctan\dfrac{\Delta_B}{\overline{R}_B} \Big)
- \overline{R}_T^3 
  \Big( \pi - 2 \arctan\dfrac{\Delta_T}{\overline{R}_T} \Big)
\bigg].
\nonumber\\
&& \hspace*{4.5cm} [\Delta_B < m_\phi, \Delta_T < m_\phi]
\end{eqnarray}
In the chiral limit, one has $\Delta_B < m_\phi$ while $\Delta_T > m_\phi$, and the corresponding NA behavior is given by
\begin{eqnarray}
\label{eq:NA-fTB-sum}
\hspace*{-1cm}
\overline{M}_{\!T} \overline{M}_{\!TB}\,
\Big(
  \Delta{\widetilde f}_{TB ({\rm on})}^{\, (1)}
+ \Delta{\widetilde f}_{TB ({\rm off})}^{\, (1)}
\Big)_{\rm NA}
&=& 
\Big[
- \tfrac{4}{3} m_\phi^2
+ \tfrac{16}{9} \big( \Delta_B^2 + \Delta_B \Delta_T + \Delta_T^2 \big)
\Big]
\log m_\phi^2
\nonumber\\
& & \hspace*{-1cm}
-\, \dfrac{16}{9 \Delta_{TB}}
\bigg[
  \overline{R}_B^3 
  \Big( \pi - 2 \arctan\dfrac{\Delta_B}{\overline{R}_B}\Big)
+ R_T^3 \log \dfrac{\Delta_T - R_T}{\Delta_T + R_T}                              
\bigg].
\nonumber\\
&& \hspace*{4.5cm} [\Delta_B < m_\phi, \Delta_T > m_\phi]
\end{eqnarray}
Finally, for the $\delta$-function contribution the LNA behavior is
\begin{equation}
\overline{M}_{\!T} \overline{M}_{\!TB}\,
\Delta{\widetilde f}_{TB (\delta)}^{\, (1)} \Big|_{\rm LNA}
= - \frac{7}{3} m_\phi^2 \log m_\phi^2.
\end{equation}
In the chiral limit, $m_\phi \to 0$, the mass difference
    $\Delta_B \sim {\cal O}(m_\phi^2)$
approaches zero first, while $\Delta_T$ remains a constant.
Further expanding
    $R_T = \Delta_T - m_\phi^2/2\Delta_T + {\cal O}(m_\phi^4)$,
the LNA behavior in Eqs.~(\ref{eq:NA-fT-sum}) and (\ref{eq:NA-fTB-sum}) can be evaluated as
\begin{eqnarray}
\overline{M}_{\!T}^2\,
\Big(
  \Delta{\widetilde f}_{T ({\rm on})}^{\, (1)}
+ \Delta{\widetilde f}_{T ({\rm off})}^{\, (1)}
\Big)_{\rm LNA}
&=& - \frac29 m_\phi^2 \log m_\phi^2,
\\
\overline{M}_{\!T} \overline{M}_{\!TB}\,
\Big(
  \Delta{\widetilde f}_{TB ({\rm on})}^{\, (1)}
+ \Delta{\widetilde f}_{TB ({\rm off})}^{\, (1)}
\Big)_{\rm LNA}
&=& + \frac43 m_\phi^2 \log m_\phi^2,
\end{eqnarray}
for the $T$ and $TB$ contributions, respectively.

Finally, combining the derived LNA behaviors for the splitting function moments with Eq.~(\ref{eq:sLNA}), the LNA contribution to the $n=1$ moment of the spin-dependent strange quark PDF in the nucleon is given by
\begin{eqnarray}
\Delta S^{(0)}_{\rm LNA} 
&=& \sum_{BT\phi} 
\frac{1}{(4\pi f_\phi)^2} 
\Big( \!
- C_{B\phi}^2\, \Delta S^{(0)}_B
+ \Delta S_{\phi (\rm tad)}^{(0)}
+ \dfrac{7}{3} C_{T\phi}^2\, \Delta S^{(0)}_T 
- C_{B\phi} C_{T\phi}\, \Delta S^{(0)}_{TB}
\Big)
m_\phi^2 \log m_\phi^2.
\nonumber\\
&&
\end{eqnarray}
Summing over all the relevant octet $B$ and decuplet $T$ states, and using the expressions for the couplings in Eqs.~(\ref{eq:C-B-phi}) and (\ref{eq:C-T-phi}) and the moments $\Delta S_h^{(0)}$ in Sec.~\ref{ssec:hadronic-pdfs}, 
we arrive at the final result for the LNA behavior of the $n=1$ strange PDF moment,
\begin{eqnarray}
\Delta S^{(0)}_{\rm LNA} &=& \frac{1}{(4\pi f_\phi)^2}
\left( \frac59 D^3 + 3 D F (D - F) + \frac12 (3F - D) \right)
m_\phi^2 \log m_\phi^2.
\end{eqnarray}
We stress that any calculation of the strange quark PDFs in the nucleon or its moments must obtain this behavior, if it is to be consistent with the chiral symmetry properties of QCD, which provides an important, model-independent constraint on nonperturbative models of the nucleon.

\section{Numerical results}

Combining the results derived in Secs.~\ref{sec:formalism} and \ref{sec:fy} for the splitting functions and the PDFs in the hadronic configurations, in this section we present the results for the numerical computation of the spin-dependent strange quark distributions in the proton.
We begin by discussing the regularization procedure for the splitting functions, and then compare the computed PDFs with some recent phenomenological parametrizations from global QCD analyses.

\subsection{Regularization of splitting functions}
\label{sec:regularization}

The hadronic splitting functions computed in Sec.~\ref{sec:fy} in the framework of chiral effective theory generally involve loop integrals that are ultraviolet divergent.
A regularization prescription is therefore required to regulate the high-energy behavior and render the loop integrals finite.
Various prescriptions have been utilized in previous analyses, including dimensional regularization~\cite{Gasser84}, finite momentum cutoffs, Pauli-Villars~\cite{plb-2016, prd-2016}, as well as finite-range regularization within local~\cite{Donoghue99, Leinweber00, Thomas03} and nonlocal~\cite{Faessler,Terning91} formulations.
Following our earlier analysis of spin-averaged strange-antistrange quark asymmetries~\cite{plb-2016, prd-2016}, we adopt here the Pauli-Villars regularization scheme, which has the advantage of preserving the Lorentz invariance, gauge invariance, and chiral symmetry of the effective theory.
It allows us to use the same phenomenological parameters as those determined in the unpolarized strange analysis~\cite{prd-2016}.

As discussed in Refs.~\cite{plb-2016, prd-2016}, the Pauli-Villars method regularizes divergent integrals by subtracting from the pointlike expressions in which the propagator masses are replaced by finite cutoff masses, such that in the high-energy limit the difference between them vanishes.
For the on-shell baryon octet splitting function, $\Delta f_B^{(\rm on)}$, we employ the subtraction
\begin{equation}\label{eq:Pauli-Villars-fH-onshell}
\frac{1}{D_\phi} = \frac{1}{k^2 - m_\phi^2} 
\to \frac{1}{k^2 - m_\phi^2} - \frac{1}{k^2 - \mu_1^2},
\end{equation}
which corresponds to using a regulating function in Eq.~(\ref{eq:fH-on}) given by
\begin{equation}
F_B^{({\rm on})}(y,k_\perp^2)
= 1 - \frac{D^2_{B\phi}}{D^2_{B\mu_1}},
\end{equation}
where $\mu_1$ is the subtraction mass parameter, and $D_{B\phi}$ is given in Eq.~(\ref{eq:D_Bphi}), and $D_{B\mu_1}$ is given by an analogous expression with $m_\phi \to \mu_1$.
A similar replacement to that in Eq.~(\ref{eq:Pauli-Villars-fH-onshell}) is made for the off-shell baryon octet function, $\Delta f_B^{({\rm off})}$, in Eq.~(\ref{eq:Deltaf_B-off}), in which case the off-shell regulating function becomes
\begin{equation}
F_B^{({\rm off})}(y,k_\perp^2)
= 1 - \frac{D_{B\phi}}{D_{B\mu_1}}.
\end{equation}
For the $\delta$-function term, $\Delta f_B^{(\delta)}$, in Eq.~(\ref{eq:fKdel}), two subtractions are necessary to take into account the divergences in both the $k^-$ and $k^2_\perp$ integrations,
\begin{equation}
\frac{1}{D_\phi}\
\to\ \frac{1}{k^2 - m_\phi^2}
   - \frac{a_1}{k^2 - \mu_1^2} 
   - \frac{a_2}{k^2 - \mu_2^2},
\end{equation}
where $\mu_1$ and $\mu_2$ are the mass parameters for the subtraction terms, whose coefficients $a_1$ and $a_2$ must satisfy the relation
\begin{eqnarray}
a_1 &=& \frac{\mu_2^2 - m_\phi^2}{\mu_2^2 - \mu_1^2}, \hspace*{1cm}
a_2  = -\frac{\mu_1^2 - m_\phi^2}{\mu_2^2 - \mu_1^2}.
\end{eqnarray}
This leads to an effective regulating function in Eq.~(\ref{eq:fKdel}) given by
\begin{equation}
F_B^{(\delta)}(y,k_\perp^2)
= 1 - \frac{a_1 \log\Omega_{\mu_1} + a_2 \log\Omega_{\mu_2}}
           {\log\Omega_\phi},
\label{eq:Fdel}
\end{equation}
with $\Omega_{\mu_i} = k_\perp^2 + \mu_i^2$.

In the decuplet sector, the loop integrals associated with the on-shell and off-shell functions are more divergent than those of the octet contributions due to the presence of derivative couplings.
Regularizing the integrals for the decuplet splitting functions, therefore, requires several subtractions, which we take to have the form
\begin{equation}
\label{eq:Pauli-Villars-decuplet-onshell}
\frac{1}{D_\phi}\
\to\ \frac{1}{k^2 - m_\phi^2} 
- \frac{b_1}{k^2 - \widetilde\mu_1^{\, 2}} 
- \frac{b_2}{k^2 - \widetilde\mu_2^{\, 2}} 
- \frac{b_3}{k^2 - \widetilde\mu_3^{\, 2}} 
- \frac{b_4}{k^2 - \widetilde\mu_4^{\, 2}},
\end{equation}
where the coefficients $b_i$ satisfy
\begin{equation}
b_i = \prod_{\stackrel{j=1}{j\neq i}}^4 
\frac{m_\phi^2 - \widetilde\mu_j^{\, 2}}
     {\widetilde\mu_i^{\, 2} - \widetilde\mu_j^{\, 2}},\ \ \ \ \ 
i=1,\ldots,4.
\end{equation}
To reduce the number of free parameters, in our numerical analysis we take $\widetilde\mu_1 = \widetilde\mu_2 = \widetilde\mu_3 = \widetilde\mu_4 \equiv \mu$ for the decuplet baryon contributions, in which case we have the replacement
\begin{equation}
\frac{1}{D_\phi}\
\to\ \frac{1}{k^2 - m_\phi^2}
\left( \frac{m_\phi^2 - \mu^2}{k^2 - \mu^2} \right)^4.
\end{equation}
For the on-shell and off-shell decuplet splitting functions in Eqs.~(\ref{eq:fT-on}) and (\ref{eq:fT-off}), the regulating functions can be written as,
\begin{eqnarray}
\label{eq:FTon}
F_T^{({\rm on})}(y, k_\bot^2)
&=& \frac{(m_\phi^2 - \mu^2)^4}{D_{T\mu}^4} 
    \bigg( 1 + \frac{4 D_{T\phi}}{D_{T\mu}} \bigg) , \\
\label{eq:FToff}
F_T^{({\rm off})}(y, k_\bot^2)
&=& \frac{(m_\phi^2 - \mu^2)^4}{D_{T\mu}^4},
\end{eqnarray}
respectively.
For the decuplet $\delta$-function contributions, Eq.~(\ref{eq:fT-del}), Pauli-Villars regularization gives the regulating functions
\begin{subequations}
\begin{eqnarray}
\hspace*{-0.5cm}
F_T^{(\delta 1)}(y,k_\bot^2)
&=& 1
- \frac{1}{\log\Omega_\phi}
  \bigg[
  \log\Omega_\mu
  + \frac{2 \Omega_\phi^3
        - 9 \Omega_\phi^2 \Omega_\mu 
        + 18 \Omega_\phi \Omega_\mu^2 
        - 11 \Omega_\mu^3}
         {6 \Omega_\mu^3} 
  \bigg],
\label{eq:Deltaf1} \\
\hspace*{-0.5cm}
F_T^{(\delta 2)}(y,k_\bot^2)
&=& 1
- \frac{1}{\Omega_\phi \log\Omega_\phi}
  \bigg[
  \Omega_\phi \log\Omega_\mu
  - \frac{\Omega_\phi^3
        - 6 \Omega_\phi^2 \Omega_\mu
        + (5 \Omega_\phi + 2 \mu^2 - 2 m_\phi^2) \Omega_\mu^2}
         {6\Omega_\mu^2}
   \bigg]
\label{eq:Deltaf2}
\end{eqnarray}
\end{subequations}
for the two functions in Eqs.~(\ref{eq:Deltaf1factor}) and (\ref{eq:Deltaf2factor}), respectively.
Finally, for the octet-decuplet transition splitting functions, the regulators in the on-shell and off-shell functions in Eqs.~(\ref{eq:fTB-on}) and (\ref{eq:fTB-off}) are given by
\begin{eqnarray}
F_{TB}^{(B)}(y, k_\bot^2) &=& \frac{(m_\phi^2-\mu^2)^4}{D_{B\mu}^4},    \\
F_{TB}^{(T)}(y, k_\bot^2) &=& \frac{(m_\phi^2-\mu^2)^4}{D_{T\mu}^4}.
\end{eqnarray}

In our previous analysis of meson loop contributions to the spin-averaged strange quark PDFs in the proton \cite{plb-2016, prd-2016}, the cutoff parameter $\mu_1$ was fixed by fitting the $p p\to \Lambda X$ differential cross section data, and an upper limit was set on $\mu_2$ by requiring that the calculated total $s + \bar{s}$ distributions do not exceed the phenomenological values, within the experimental uncertainties, for any value of $x$.
The best fit gave $\{\mu_1, \mu_2\} = \{545, 600\}$\,MeV, while the set $\{\mu_1, \mu_2\} = \{526, 894\}$\,MeV resulted in two standard deviations below the best fit. 
For the cutoff parameter $\mu$ in the decuplet sector, a good fit to the $p p \to \Sigma^{*+} X$ differential cross section data~\cite{BHM:1978} was achieved with $\mu = 762(21)$~MeV.
In the present analysis of spin-dependent PDFs we use the same parameters, along with SU(3) symmetric values of the couplings $C_{B\phi}$ and $C_{T\phi}$, to compute the splitting functions numerically.

\begin{figure}[t]
\includegraphics[width=\columnwidth]{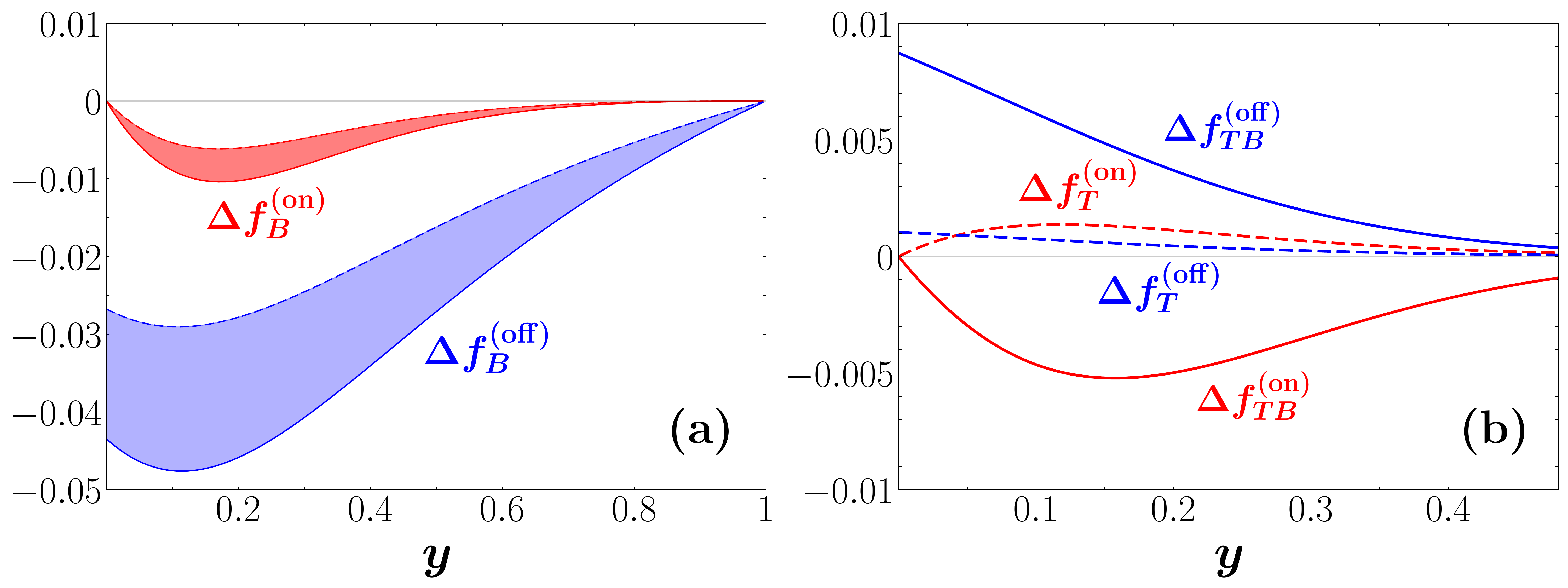}
\caption{Momentum dependence of the spin-dependent splitting functions for {\bf (a)} octet baryon $\Delta f_B$ and {\bf (b)} decuplet baryon $\Delta f_T$ (dashed lines) and octet-decuplet interference $\Delta f_{TB}$ (solid lines) intermediate states for the on-shell (red bands and curves) and off-shell (blue bands and curves) contributions.
The octet results are computed for the $\Sigma^0 K^+$ intermediate state with the cutoffs in the range $\{\mu_1,\mu_2\} = \{545,600\}$ to $\{526,894\}$\,MeV for the upper (dashed) and lower (solid) edges of the bands, respectively, while the decuplet results are computed for the $\Sigma^{*0} \Sigma^0 K^+$ intermediate state with a cutoff $\mu=762$\,MeV.}
\label{f:splitting-functions}
\end{figure}

The spin-dependent splitting functions for the strange octet, decuplet and octet-decuplet baryon interference intermediate states are shown in Fig.~\ref{f:splitting-functions}, for the on-shell and off-shell contributions.
For the octet baryon splitting functions [Fig.~\ref{f:splitting-functions}(a)], both the on-shell $\Delta f_B^{(\rm on)}$ and off-shell $\Delta f_B^{(\rm off)}$ polarized functions are negative for all values of $y$, peaking at $y \approx 0.1-0.2$.
Interestingly, the off-shell function has a magnitude that is several times larger than the on-shell function.
Compared with the analogous spin-averaged results \cite{prd-2016}, the (negative) spin-dependent on-shell function is about 4--5 times smaller in magnitude, while the off-shell function is identical in both cases
(there is a small difference arising from the different baryon masses between $\Lambda$ and $\Sigma^0$).
The uncertainties on the on-shell and off-shell distributions arising from the choice of cutoffs $\mu_1$ and $\mu_2$, indicated by the bands, is smaller than the difference between the respective on-shell and off-shell results.

For the splitting functions that involve decuplet baryons in the intermediate state [Fig.~\ref{f:splitting-functions}(b)], the on-shell contributions vanish at $y=0$, while the off-shell contributions remain nonzero. 
The decuplet on-shell and off-shell splitting functions are both positive, while there is strong cancellation between these two pieces for the octet-decuplet interference splitting function.
Note that since $\Delta f_{TB}^{(\rm on)}$ and $\Delta f_{TB}^{(\rm off)}$ are multiplied by the couplings $C_{T\phi} C_{B\phi}$ in Eq.~(\ref{eq:f-TBK}), which for the $\Sigma^{* 0} \Sigma^0 K^+$ case is negative [Eqs.~(\ref{eq:C-T-phi}) and (\ref{eq:C-B-phi})], the sign of the overall contribution of these terms can be opposite to that shown in Fig.~\ref{f:splitting-functions}.

\subsection{Polarized strange quark distributions}
\label{sec:results}

With the hadronic splitting functions thus determined, the remaining ingredients needed to proceed with the evaluation of the polarized strange quark PDF in the proton are the PDFs in the hadronic configurations in Sec.~\ref{ssec:hadronic-pdfs}.
Specifically, the SU(3) relations in Eqs.~(\ref{eq:slambda-ssigma})--(\ref{eq:s-KR}), (\ref{eq:sSig*}) and (\ref{eq:sSig*Sig}) connect the strange quark PDFs for the various intermediate states with the spin-dependent and spin-averaged $u$ and $d$ quark PDFs in the proton.
The PDFs in the proton are relatively well determined from global analyses of high-energy polarized~\cite{JAM17, NNPDF:2014, Hartland:2013} and unpolarized~\cite{Ball:2017nwa, JAM19} cross section data.
For the spin-averaged $u$ and $d$ quark distributions in the proton, for convenience we use the recent CJ15 parametrization~\cite{Accardi:2016} at $Q^2=1$~GeV$^2$, while the polarized PDFs, $\Delta u$ and $\Delta d$, are taken from the JAM analysis~\cite{JAM17} at the same scale.
We have also performed the analysis with other unpolarized~\cite{MRST:1998} and polarized~\cite{LSS15} PDF sets, and found the dependence on the choice of input parametrization relatively mild.

For representing the contributions to the polarized strange PDF from the various terms in Eq.~(\ref{eq:convolution-2}), it is convenient to express the total distribution in terms of the diagrams in Fig.~\ref{fig:loop-octet-decuplet}.
Decomposing each diagram into on-shell, off-shell and $\delta$-function contributions, in analogy with the unpolarized case in Ref.~\cite{prd-2016}, one can write the total $\Delta s$ PDF as
\begin{subequations}
\label{eq:decomposition}
\begin{eqnarray}
\label{eq:decomposition_diag}
\Delta s(x)
&=& \big( \Delta s^{\rm (on)} + \Delta s^{\rm (off)} + \Delta s^{(\delta)} \big)_{B\, \rm rbw}\
 +\ \big( \Delta s^{\rm (off)} + \Delta s^{(\delta)} \big)_{\rm KR}\
 +\ \big( \Delta s^{(\delta)} \big)_{\rm tad}\
    \nonumber\\
& & 
 +\ \big( \Delta s^{\rm (on)} + \Delta s^{\rm (off)} + \Delta s^{(\delta)} \big)_{T\, \rm rbw}\
 +\ \big( \Delta s^{\rm (on)} + \Delta s^{\rm (off)} + \Delta s^{(\delta)} \big)_{TB\, \rm rbw}\
    \\
&=& \underbrace{\Delta s^{\rm (on)}_{B\, \rm rbw}
              + \Delta s^{\rm (on)}_{T\, \rm rbw}
              + \Delta s^{\rm (on)}_{TB\, \rm rbw}}_{\rm on-shell}\
 +\ \underbrace{\Delta s^{\rm (off)}_{B\, \rm rbw}
              + \Delta s^{\rm (off)}_{T\, \rm rbw}
              + \Delta s^{\rm (off)}_{TB\, \rm rbw}
              + \Delta s^{\rm (off)}_{\rm KR}}_{\rm off-shell}
    \nonumber\\
& &
 +\ \underbrace{\Delta s^{(\delta)}_{B\, \rm rbw}
              + \Delta s^{(\delta)}_{T\, \rm rbw}
              + \Delta s^{(\delta)}_{TB\, \rm rbw}
		      + \Delta s^{(\delta)}_{\rm KR}
		      + \Delta s^{(\delta)}_{\rm tad}}_{\rm \delta-function}.
\label{eq:decomposition_type}
\end{eqnarray}
\end{subequations}
Note that the on-shell contributions arise only from the (octet, decuplet and octet-decuplet interference) baryon rainbow diagrams [Figs.~\ref{fig:loop-octet-decuplet}(a),~\ref{fig:loop-octet-decuplet}(d), and~\ref{fig:loop-octet-decuplet}(e)], the off-shell terms come from rainbow and Kroll-Ruderman diagrams [Fig.~\ref{fig:loop-octet-decuplet}(c)], while all diagrams, including the tadpole [Fig.~\ref{fig:loop-octet-decuplet}(b)], contribute to the $\delta$-function terms.

\begin{figure}[t]
\includegraphics[width=\columnwidth]{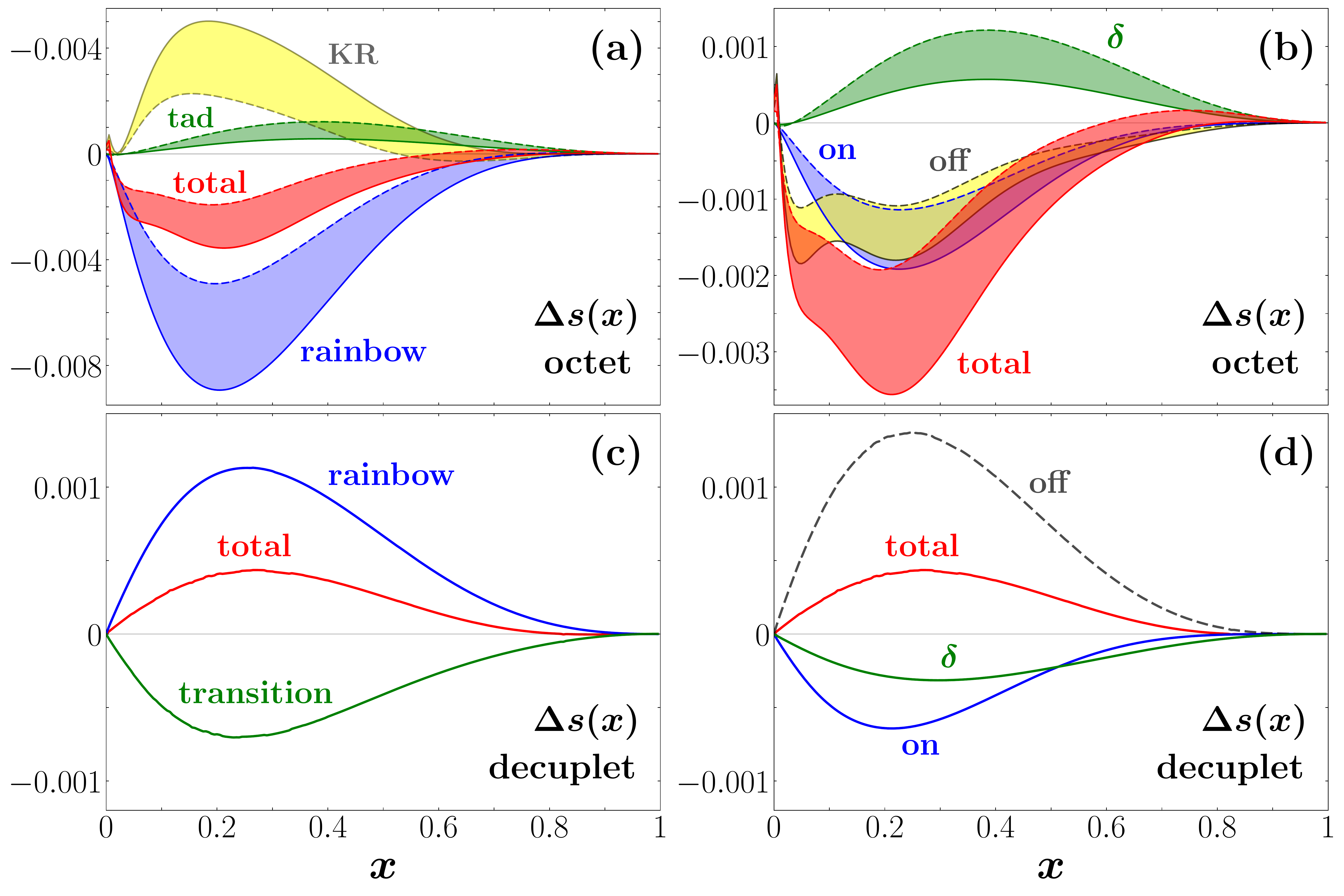}
\caption{Contributions to the $x\Delta s$ distribution in the proton at $Q^2=1$~GeV$^2$ from various meson loop diagrams with octet intermediate states [panels (a) and (b)] and decuplet (and decuplet-octet interference) states [panels (c) and (d)]. The bands for the octet contributions correspond to the range of parameters $\{\mu_1,\mu_2\} = \{545,600\}$\,MeV to $\{526,894\}$\,MeV for the dashed and solid edges of the bands, respectively, while the decuplet results use $\mu=762$\,MeV.  The left column [panels (a) and (c)] corresponds to the decomposition according to the diagram type [Fig.~\ref{fig:loop-octet-decuplet} and Eq.~(\ref{eq:decomposition_diag})], while the right column [panels (b) and (d)] corresponds to the decomposition according to the function type [Eq.~(\ref{eq:decomposition_type})].}
\label{fig:xs_decomp}
\end{figure}

The contributions to the polarized strange PDF $x\Delta s$ from the various terms in Eqs.~(\ref{eq:decomposition}) are shown in Fig.~\ref{fig:xs_decomp}, for both the decompositions in terms of types of diagrams [Eq.~(\ref{eq:decomposition_diag})] and types of functions [Eq.~(\ref{eq:decomposition_type})].
For the octet baryon states, we find [Fig.~\ref{fig:xs_decomp}(a)] large cancellations between the negative rainbow and positive KR diagrams, with the tadpole diagram making a smaller and positive contribution.
The result is a negative total octet baryon contribution to $x \Delta s$ that is about 1/3 of the size of the rainbow, peaking at $x \approx 0.2$.

A somewhat clearer picture of the cancellations is revealed when we look at the total on-shell, off-shell, and $\delta$-function contributions in Fig.~\ref{fig:xs_decomp}(b) from all octet baryon diagrams.
At intermediate values of $x$, the negative on-shell and off-shell components give comparable contributions, with the off-shell dominating at smaller~$x$.
In contrast, the $\delta$-function piece is positive, with a broad shape peaking at $x \sim 0.3-0.4$.
Its overall magnitude is smaller than the other contributions, so that it only partially cancels the negative on-shell and off-shell terms, leaving the total $x \Delta s$ distribution peaking at around $-0.002$ to $-0.003$ for $x \sim 0.2$.

For the diagrams involving intermediate states with decuplet baryons, shown in Figs.~\ref{fig:xs_decomp}(c) and~\ref{fig:xs_decomp}(d), there are again large cancellations between positive decuplet rainbow and negative octet-decuplet transition contributions, whose overall magnitude is smaller than those from the octet states.
Furthermore, in contrast to the octet case, the off-shell contribution is positive, but canceled somewhat by the negative on-shell and $\delta$-function terms, which turn out to have a very similar shape and magnitude.
The net result is a total positive effect, with about 1/5 of the magnitude of the octet contribution.

\begin{figure}[t]
\includegraphics[width=0.8\columnwidth]{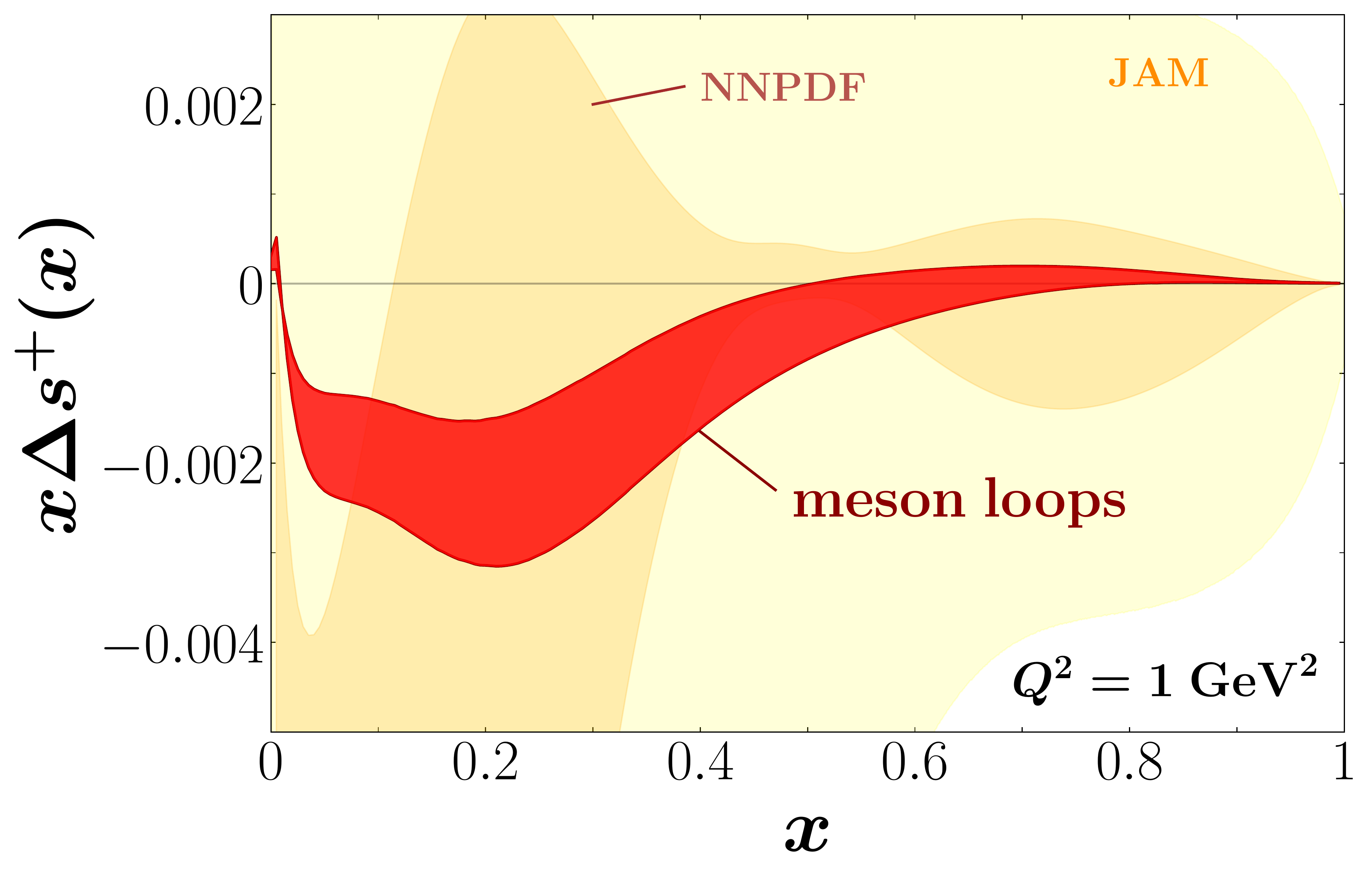}
\caption{Comparison of the calculated total meson loop contribution to the polarized strange quark PDF (dark red band) with $x \Delta s^+ \equiv x\Delta s + x\Delta \bar s$ from the phenomenological NNPDF~\cite{NNPDF:2014, Hartland:2013} (orange band) and JAM~\cite{JAM17} (yellow band, spanning most of the graph) global QCD analyses at $Q^2 = 1$~GeV$^2$.
The band for the meson loop contributions corresponds to the range of cutoff parameters $\{\mu_1,\mu_2\} = \{545,600\}$~MeV to $\{526,894\}$~MeV for octet baryons and $\mu=762$~MeV for decuplet baryons.}
\label{f.PDFs}
\end{figure}

Comparing the calculated polarized strange distribution with phenomenological PDFs obtained from global QCD analyses, in Fig.~\ref{f.PDFs} we show the total $x\Delta s$ from the chiral theory together with parametrizations from the NNPDF~\cite{NNPDF:2014} and JAM~\cite{JAM17} analyses at $Q^2=1$~GeV$^2$.
The most striking observation is the small magnitude of the calculated strange polarization compared with the uncertainty bands of the global parametrizations, which reflects the relatively weak constraints on $\Delta s$ that exist from current experiments.
The JAM study~\cite{JAM17}, in particular, performed a dedicated analysis of the strange quark PDF using data from inclusive and semi-inclusive DIS, without imposing the commonly used assumption about SU(3) flavor symmetry for the axial charges extracted from hyperon decays~\cite{Bass10}.
This leads to a significantly larger uncertainty on $\Delta s$ than that obtained in analyses that do impose SU(3) symmetry on the axial charges~\cite{DSSV09, DSSV14, AAC09, BB10, LSS10, LSS11, LSS15, KTA17, NNPDF:2014, JAM15}.

Furthermore, since existing data cannot discriminate between the strange quark and antiquark polarizations, all of the global QCD analyses assume that
    $\Delta s = \Delta \bar s$, 
so that in practice 
    $\Delta s^+ \equiv \Delta s + \Delta \bar s \to 2 \Delta s$.
In contrast, in the chiral theory calculation, assuming valence dominance of the bare hadronic state wave functions, the only source of strangeness in the proton is the coupling to the strange meson--baryon intermediate states.
Since all strange antiquarks reside in the spin-0 kaon, in this framework the antistrange polarization $\Delta \bar s$ is identically zero.
One may therefore expect the determinations of the strange polarization in the global QCD analyses to overestimate the $\Delta s$ contribution from the chiral calculation.

\begin{table}[tb]
\begin{center}
\caption{\label{tab:moment-octet}
Individual contributions to the first moment of $\Delta s(x)$ at $Q^2 = 1$~GeV$^2$, in units of $10^{-2}$, summed over the appropriate octet and decuplet hyperon states. The contributions from octet, decuplet and octet-decuplet interference intermediate states, as in Eq.~(\ref{eq:decomposition_type})  are listed separately. The sum of all contributions to the total moment is in the range
    $\bm{\langle \Delta s \rangle = [-0.50,-0.25] \times 10^{-2}}$.\\}
\begin{tabular}{c|cccccc|c}     \hline\hline
~$\{\mu_1,\mu_2\}$\,(MeV)                            &
~$\langle \Delta s \rangle_{B\, \rm rbw}^{(\rm on)}$ &
~$\langle \Delta s \rangle_{B\, \rm rbw}^{(\rm off)}$&
~$\langle \Delta s \rangle_{B\, \rm rbw}^{(\delta)}$ &
~$\langle \Delta s \rangle_{\rm KR}^{(\rm off)}$     &
~$\langle \Delta s \rangle_{\rm KR}^{(\delta)}$      &
~$\langle \Delta s \rangle_{\rm tad}^{(\delta)}$~    &
~total~ \\ \hline
\{545, 600\}   &
$-0.40$ &
$-1.62$ &
~~0.07  &
~~1.43  &
$-0.15$ &
~~0.08  &   
~$-0.59$~     \\ 
\{526, 894\}   &
$-0.23$ &
$-0.98$ &
~~0.15  &
~~0.86  &
$-0.31$ &
~~0.17  &   
~$-0.34$~     \\ \hline\hline
~$\mu$\,(MeV)                                   &
~$\langle \Delta s \rangle_{T\, \rm rbw}^{(\rm on)}$   &
~$\langle \Delta s \rangle_{T\, \rm rbw}^{(\rm off)}$  &
~$\langle \Delta s \rangle_{T\, \rm rbw}^{(\delta)}$   &
~$\langle \Delta s \rangle_{TB\, \rm rbw}^{(\rm on)}$  &
~$\langle \Delta s \rangle_{TB\, \rm rbw}^{(\rm off)}$  &
~$\langle \Delta s \rangle_{TB\, \rm rbw}^{(\delta)}$~ &
~total~ \\ \hline
762     &
~~0.10 &
~~0.05  &
~~0.10  &
$-0.25$  &
~~0.26 &
$-0.17$ &   
~$+0.09$~     \\ \hline\hline
\end{tabular}
\end{center}
\end{table}

Integrating the calculated distribution over all $x$, in Table~\ref{tab:moment-octet} we list the contributions of the various terms in Eq.~(\ref{eq:decomposition_type}) to the lowest ($n=1$) moment of $\Delta s(x)$, which from Eq.~(\ref{eq:mom_def}) we denote by
    $\langle x^0 \rangle_{\Delta s} \equiv \langle \Delta s \rangle$.
Numerically, a large degree of cancellation is seen between the various on-shell and off-shell terms, with the $\delta$-function terms being somewhat smaller.
Within the range of cutoff parameters considered in this analysis, the octet baryon intermediate state contributions to $\langle \Delta s \rangle$ are found to be in the range $-0.006$ to $-0.003$, while the contribution from decuplet baryon intermediate states is $\approx +0.003$ and from octet-decuplet interference $\approx -0.002$.
The net polarization in the proton carried by strange quarks is then predicted to be in the range $\langle \Delta s \rangle \approx [-0.0050, -0.0025]$ within the uncertainties of the cutoff parameters.

This can be compared with the value determined from the JAM global QCD analysis~\cite{JAM17} of 
    $\langle \Delta s^+ \rangle_{\scriptsize{\rm JAM}} = - 0.03(10)$.
While our central values are about an order of magnitude smaller than the phenomenological results, they are in good agreement within the relatively large uncertainty.
Future data on semi-inclusive DIS and parity-violating inclusive DIS from the planned Electron-Ion Collider~\cite{EIC} should reduce the uncertainty on the extracted $\langle \Delta s^+ \rangle$ and allow a better discrimination between the $\Delta s$ and $\Delta \bar s$ distributions.

\section{Conclusion}
\label{sec:conclusion}

In summary, we have performed a comprehensive study of the polarized strange quark distribution in the proton within chiral effective field theory at the one meson loop level.
The full set of spin-dependent proton $\to$ meson $+$ baryon splitting functions was computed, including contributions from octet and decuplet rainbow diagrams, as well as tadpole, Kroll-Ruderman and octet-decuplet transition diagrams.
From these we derived the leading nonanalytic behavior of the lowest moment of the polarized strange quark PDF, finding the characteristic $m_\phi^2 \log m_\phi^2$ form with a coefficient depending on low-energy baryon~properties.

We have used the Pauli-Villars regularization scheme to regularize the ultraviolet divergences in the loop integrals, with cutoff parameters determined from comparison of the spin-averaged distributions with semi-inclusive hyperon production in $pp$ collisions.
With these parameters the octet intermediate state contributions are dominated by the negative on-shell term, with further enhancement from the off-shell term at low $x$, and partial cancellation from the positive $\delta$-function component.
Some cancellation also exists between the positive decuplet rainbow and the negative octet-decuplet contributions, with both on-shell and off-shell terms playing an important role.

The result is that the octet contributions are mostly responsible for the polarized strange PDF $\Delta s(x)$ being negative at small $x$, with the lowest moment, $\langle \Delta s \rangle$, lying in the range $(-5.0, -2.5) \times 10^{-3}$. 
In comparison with the recent JAM global QCD analysis,
    $\langle \Delta s^+ \rangle_{\scriptsize{\rm JAM}} = -0.03(10)$~\cite{JAM17},
or the latest lattice QCD calculation from the ETM Collaboration,
    $\langle \Delta s^+ \rangle_{\scriptsize{\rm latt}} = -0.046(8)$~\cite{Alexandrou:2020sml},
the chiral contribution is relatively small, although consistent with the phenomenological values within the uncertainties.

In the future it will be important to compare the current work with calculations within a nonlocal chiral theory, such as that used for the unpolarized sea quark asymmetries in Refs.~\cite{Salamu:2019-1, Salamu:2019-2}.
Furthermore, extending the analysis to the nonstrange (valence quark) distributions $\Delta u(x)$ and $\Delta d(x)$ using the relativistic formalism presented here should provide robust estimates of the effect of the chiral effects on the axial charges $g_A$ and $g_8$ and total helicity $\Delta \Sigma$ carried by quarks.

\section*{Acknowledgments}

This work is supported by the Australian Research Council through the ARC Centre of Excellence for Particle Physics at the Terascale (CE110001104) and Discovery Projects DP151103101 and DP180100497 (AWT), the DOE Contract No.~DE-AC05-06OR23177, under which Jefferson Science Associates, LLC operates Jefferson Lab, DOE Contract No.~DE-FG02-03ER41260, and by the NSFC under Grant No.~11975241.

\appendix
\section{Derivation of decuplet and octet-decuplet splitting functions}
\label{sec:fT}

In this appendix we present some details about the derivation of the decuplet rainbow splitting function $\Delta f_{T\phi}^{(\rm rbw)}$ in Eqs.~(\ref{eq:decup-rbw})--(\ref{eq:NiT}) and the octet-decuplet transition splitting function $\Delta f_{TB\phi}^{(\rm rbw)}$ in Eqs.~(\ref{eq:f-TB})--(\ref{eq:NiTB}) using the Pauli-Villars regularization scheme as discussed in Sec.~\ref{sec:regularization}.
After performing $k^-$ integration in Eq.~(\ref{eq:decup-rbw}), the first term gives rise to
\begin{eqnarray}
\label{eq:fT-N1}
&& - \frac{i}{2M s^+} \frac{C_{T\phi}^2}{f_\phi^2}
\int\!\frac{d^4 k}{(2\pi)^4} 
\frac{N_1^T}{D^2_T D_\phi}
\frac{(m_\phi^2 - \mu^2 )^4}{(k^2 - \mu^2)^4}
\delta\Big(y-\frac{k^+}{p^+} \Big)
\nonumber\\
&=& -
\frac{C_{T\phi}^2}{(4\pi f_\phi)^2} 
\frac{(m_\phi^2 - \mu^2)^4}{(3 M_T)^2} 
\int dk^2_\bot
\nonumber\\ 
&& \hspace*{1.5cm} \times
\Bigg\{
\frac{y \big[ k_\bot^2 + (M_T + \yb M)^2 \big] 
        \big[ k_\bot^4 - 8 \yb M M_T k_\bot^2 - \big(M_T^2 - \yb^2 M^2\big)^2 \big]}
     {2 \yb^4 D_{T\phi}^2 D_{T\mu}^4} 
\Big( 1 + \frac{4 D_{T\phi}}{D_{T\mu}} \Big)
\nonumber\\ 
&& \hspace*{2cm}
+ \frac{k_\bot^4 - 5 \yb M M_T k_\bot^2
        - \big( M_T + \yb M \big)^2 \big( M_T^2 + \yb M M_T + \yb^2 M^2 \big)}
       {\yb^3 D_{T\phi} D_{T\mu}^4} 
\Bigg\},
\end{eqnarray}
where $D_{T\phi}$ is given by Eq.~(\ref{eq:D_Tphi}), and $D_{T\mu}$ is given by an analogous expression with $m_\phi \to \mu$.
The second term in Eq.~(\ref{eq:decup-rbw}) can be written as
\begin{eqnarray}
\label{eq:fT-N2}
&& - \frac{i}{2M s^+}
\frac{C_{T\phi}^2}{f_\phi^2}
\int\!\frac{d^4 k}{(2\pi)^4} \frac{N_2^T}{D_T D_\phi}
\frac{(m_\phi^2 - \mu^2)^4}{(k^2 - \mu^2)^4}
\delta\Big(y-\frac{k^+}{p^+}\Big)
\nonumber\\
&=& \frac{C_{T\phi}^2}{(4\pi f_\phi)^2}
\frac{(m_\phi^2 - \mu^2)^4}{(3 M_T^2)^2}
\int dk^2_\bot
\dfrac{1}{\yb^3 D_{T\phi} D_{T\mu}^4}
\bigg\{
k_\bot^6 - \yb \big[ 3 M M_T - \yb M^2 \big] k_\bot^4
\nonumber\\
&& \hspace*{3.5cm}
- \big[ 3 M_T^4 + 7 \yb M M_T^3 + 4 \yb^2 M^2 M_T^2 + 6 \yb^3 M^3 M_T + \yb^4 M^4 \big] k_\bot^2
\nonumber\\
&& \hspace*{3.5cm}
- \big[ 2 M_T^4 - \yb^2 M^2 M_T^2 + \yb^3 M^3 M_T + \yb^4 M^4 \big] 
  \big( M_T + \yb M \big)^2
\bigg\}.
\end{eqnarray}
The term proportional to $1/D_{T\phi}^2$ in Eq.~(\ref{eq:fT-N1}) is identified as the on-shell splitting function, consistent with the result in Ref.~\cite{Holtmann96}, which gives rise to Eq.~(\ref{eq:fT-on}) and the regulating function in Eq.~(\ref{eq:FTon}). 
The sum of the terms proportional to $1/D_{T\phi}$ in Eqs.~(\ref{eq:fT-N1}) and (\ref{eq:fT-N2}) gives rise to the decuplet baryon off-shell function in Eqs.~(\ref{eq:fT-off}) and (\ref{eq:FToff}).
Finally, the $1/D_{T\phi}$ term in Eq.~(\ref{eq:decup-rbw}) that gives rise to the $\delta$-function term involves the integral,
\begin{eqnarray}
&& 
\nonumber\\
&& \int d^4 k \frac{1}{D_\phi}
\frac{(m_\phi^2 - \mu^2)^4}{(k^2 - \mu^2)^4}\,
\delta\Big(y-\frac{k^+}{p^+}\Big)
\nonumber\\
&=& 4 (m_\phi^2 - \mu^2)^4 \int d^4k
\int_0^1 dz
\frac{z^3}{ \big[ z(k^2 - \mu^2 + i\epsilon) + (1-z)(k^2 - m_\phi^2 + i \epsilon)
            \big]^5}\,
\delta\Big(y-\frac{k^+}{p^+}\Big)
\nonumber\\
&=& \frac16
\frac{\partial^4}{\partial \Omega^4}
\int_0^1 dz\, z^3
\int d^4k \frac{1}{(k^2 - \Omega + i \epsilon)}\,
\delta\Big(y-\frac{k^+}{p^+}\Big)
\nonumber\\
&=& \frac{i \pi^2}{6} 
\frac{\partial^4}{\partial \Omega^4}
\int_0^1 dz\, z^3 
\int dk^2_\perp\, \log(k^2_\perp + \Omega)\,
\delta(y)
\\
&=& - i \pi^2 \int dk^2_\perp
\int_0^1 dz\, \frac{z^3}{(k^2_\perp + \Omega)^4}\,
\delta(y)
\nonumber\\
&=& i \pi^2 \int dk^2_\perp 
\bigg[ 
\log\frac{\Omega_\phi}{\Omega_\mu}
- \frac{2 \Omega_\phi^3 - 9 \Omega_\phi^2 \Omega_\mu + 18 \Omega_\phi \Omega_\mu^2}
       {6 \Omega_\mu^3} + \frac{11}{6}
\bigg] \delta(y),
\nonumber\\
&&
\nonumber
\end{eqnarray}
where 
\begin{eqnarray}
\Omega      &=& z \mu^2 + (1-z) m_\phi^2,   \nonumber\\
\Omega_\phi &=& k^2_\perp + m^2_\phi,       \\
\Omega_\mu  &=& k^2_\perp + \mu^2.          \nonumber
\end{eqnarray}
Similarly, we can compute the integral
\begin{eqnarray}
\label{eq:A5}
&&
\nonumber\\
&&\int d^4k\, \frac{2 y\, p \cdot k}{D_\phi}
\frac{(m^2_\phi - \mu^2)^4}{(k^2 - \mu^2)^4}\,
\delta\Big(y-\frac{k^+}{p^+}\Big)
\nonumber\\
&=& i \pi^2 \int dk^2_\perp
\bigg[ 
    \Omega_\phi \Big( \log\frac{\Omega_\phi}{\Omega_\mu} + \frac56 \Big)
    - \frac13 (m^2_\phi - \mu^2) + \frac{\Omega_\phi^3}{6 \Omega_\mu^2}
    - \frac{\Omega_\phi^2}{\Omega_\mu}
\bigg]\, \delta(y).
\\
&&
\nonumber
\end{eqnarray}
Combining the results in Eqs.~(\ref{eq:fT-N1})--(\ref{eq:A5}), we then arrive at the expressions for the on-shell, off-shell and $\delta$-function decuplet splitting functions in Eqs.~(\ref{eq:fT-on}), (\ref{eq:fT-off}) and (\ref{eq:fT-del}), respectively.

\newpage

For the octet-decuplet transition splitting function $\Delta f_{TB\phi}^{(\rm rbw)}$, following the same procedure we have for the first term in Eq.~(\ref{eq:f-TB}),
\begin{eqnarray}
\label{eq:fTB-N1}
&& \frac{i}{2M s^+}
\frac{C_{T\phi} C_{B\phi}}{f_\phi^2} 
\int\!\frac{d^4 k}{(2\pi)^4}
\frac{N_1^{TB}}{D_T D_B D_\phi}
\frac{(m_\phi^2 - \mu^2)^4}{(k^2 - \mu^2)^4}
\delta\Big(y - \frac{k^+}{p^+}\Big)
\nonumber\\
&=&
\frac{C_{T\phi} C_{B\phi} \overline{M}_{\!T} \overline{M}_{\!TB} }
     {(4\pi f_\phi)^2} 
\frac{(m_\phi^2 - \mu^2)^4}{3 M_T^2 \overline{M}_{\!TB} \Delta_{TB}}
\int dk_\bot^2
\nonumber\\
&& \times 
\bigg\{
\frac{1}{\yb^2}
\Big[ \frac{1}{D_{T\phi} D_{T\mu}^4} - \frac{1}{D_{B\phi} D_{B\mu}^4} \Big]
\nonumber\\
&& \hspace*{0.8cm} \times
\bigg[ 
k_\bot^4
- \big( 2 M_T \Delta_{TB} + \yb M (3 M_T - M_B) \big) k_\bot^2
- \big(\Delta_B + y M\big) \big(\Delta_T + y M \big) \big(\overline{M}_{\!T} - y M\big)^2
\bigg]
\nonumber\\
&& \hspace*{0.5cm}
+\, \frac{\Delta_{TB}}{\yb^2 D_{T\phi} D_{T\mu}^4}
\Big[ \big(2 M_T + \yb M\big) k_\bot^2 
    - \big(\Delta_T + y M\big) \big(\overline{M}_{\!T} - y M\big)^2
\Big] 
\nonumber\\
&& \hspace*{0.5cm}
-\, \frac{\Delta_{TB}}{\yb^2 D_{B\phi} D_{B\mu}^4}
\Big[ \big(\Delta_{TB} + \yb M\big) k_\bot^2
    + \yb^2 M^2 \big(2 M_T + \yb M\big)
    + M_B \big(M_T^2 + \yb M M_T + \yb^2 M^2 \big)
\Big]
\bigg\}.
\nonumber\\
&&
\end{eqnarray}
For the second term in Eq.~(\ref{eq:f-TB}), we can write
\begin{eqnarray}
\label{eq:fTB-N2}
&& \frac{i}{2M s^+} \frac{C_{T\phi} C_{B\phi}}{f_\phi^2} 
\int\!\frac{d^4 k}{(2\pi)^4}
\frac{N_2^{TB}}{D_B D_\phi}
\frac{(m_\phi^2 - \mu^2)^4}{(k^2 - \mu^2)^4}
\delta\Big(y - \frac{k^+}{p^+}\Big)
\nonumber\\
&=&
\frac{C_{T\phi} C_{B\phi} \overline{M}_{\!T} \overline{M}_{\!TB}}
     {(4\pi f_\phi)^2}  
\frac{(m_\phi^2 - \mu^2)^4}{3 M_T^2 \overline{M}_{\!T} \overline{M}_{\!TB}} \int dk^2_\bot\,
\frac{1}{\yb^2 D_{B\phi} D_{B\mu}^4} 
\nonumber\\
&& \times 
\bigg\{
k_\bot^4 
- \Big[ M_T^2 
    - 2 M_T \overline{M}_B
    + M M_B 
    - \yb \big(4 M^2 + M M_T + 3 M_T M_B + 4 M M_B \big)
  \Big] k_\bot^2
\nonumber\\
&& \hspace*{0.5cm}
-\, M_B^4
- M_B^3 \big(\overline{M}_{\!T} - 3 y M_T\big)
- M_B^2 \big[ M_T^2 - 2 \yb^2 M^2 + (1 + 3y - 6 y^2) M M_T \big]
\nonumber\\
&& \hspace*{0.5cm}
+\, \yb M
  \big[ M_T^2 \overline{M}_{\!T} + 3 \yb M M_T^2 + 3 \yb^2 M^2 M_T - \yb^3 M^3 \big]
\nonumber\\
&& \hspace*{0.5cm}
+\, \yb M M_B \big[ M_T^2 + (1 + 3 \yb^2) M M_T + 3 \yb M^2 \big]
\bigg\}.
\end{eqnarray}
As for the decuplet rainbow diagram, the first term in the braces of Eq.~(\ref{eq:fTB-N1}) is defined as the on-shell octet-decuplet splitting function, Eq.~(\ref{eq:fTB-on}), consistent with the result of Ref.~\cite{Holtmann96}, and the remaining part is combined with Eq.~(\ref{eq:fTB-N1}) to give to the off-shell octet-decuplet splitting function, Eq.~(\ref{eq:fTB-off}).

\newpage

\end{document}